\documentclass[reprint, amsmath,amssymb, aps, nofootinbib]{revtex4-1}

\usepackage{natbib}
\usepackage{graphicx}
\usepackage{dcolumn}
\usepackage{bm}
\preprint{APS/123-QED}

\makeatletter
\newcommand*{\rom}[1]{\expandafter\@slowromancap\romannumeral #1@}
\makeatother

\begin{document}

\title{Global and Local diffusion\\in the Standard
Map}

\author{Mirella Harsoula}
 \email{mharsoul@academyofathens.gr}
\author{George Contopoulos}%
 \email{gcontop@academyofathens.gr}
\affiliation{%
 Research Center for Astronomy,
           Academy of Athens\\ Soranou Efesiou 4, GR-115 27 Athens, Greece}

\date{\today}

\begin{abstract}
We study the global and the local transport and diffusion in the case of the standard map, by calculating the diffusion exponent $\mu$. In the global case we find that the mean diffusion exponent for the whole phase space is either $\mu=1$, denoting normal diffusion or  $\mu=2$ denoting anomalous diffusion (and ballistic motion). The mean diffusion of the whole phase space is normal when no accelerator mode exist and it is anomalous  (ballistic) when accelerator mode islands exist even if their area is tiny in the phase space. The local value of the diffusion exponent inside the normal islands of stability is $\mu=0$, while inside the accelerator mode islands it is $\mu=2$. The local value of the diffusion exponent in the chaotic region outside the islands of stability converges always to the value of 1.  The time of convergence can be very long, depending on the  distance from the accelerator mode islands and the value of the non linearity parameter $K$. For some values of $K$ the stickiness around the accelerator mode islands is maximum and initial conditions inside the sticky region can be dragged in a ballistic motion for extremely long times of the order of $10^7$ or more but they will finally end up in normal mode diffusion with $\mu=1$. We study, in particular, cases with maximum stickiness and cases where normal and accelerator mode islands coexist. We find general analytical solutions of periodic orbits of accelerator type and we give evidence that they are much more numerous than the normal periodic orbits. Thus, we expect that in every small interval $\Delta K$ of the non linearity parameter $K$ of the standard map there exist smaller intervals of accelerator mode islands. However, these smaller intervals are in general very small, so that in the majority of the values of $K$ the global diffusion is normal.   

\begin{description}
\item[PACS numbers]
05.45.-a
\end{description}
\end{abstract}

\maketitle

\section{Introduction}

A problem of great interest in  dynamical
systems is the diffusion of the orbits. There are two main types of diffusion, normal and anomalous. In general the anomalous diffusion is much faster than the normal diffusion, and it has important consequences. Several people have worked on this subject (\cite{chirikov1979universal},\cite{israilev1990simple}, \cite{ma1992lieberman}, \cite{zumofen1994random},\cite{1995LNP...450..196K}, \cite{afraimovich1997fractal}, \cite{1999PhRvE..59...3756Z},\cite{2007PhRvE..75c6213K}, \cite{venegeroles2008calculation},  \cite{manos2014survey}, \cite{meiss2015thirty}). One of the  most simple cases of 2-D mappings is the standard map:

\begin{eqnarray}\label{stand}
\nonumber \hspace{1cm} y' = y +\frac{K}{2 \pi} sin (2 \pi x )
\end{eqnarray}
\begin{eqnarray}
\hspace{1cm} x' = x + y' \hspace{2cm} (mod 1)
\end{eqnarray}
assuming that the modulo applies only on the x-coordinate and the diffusion takes place in the y-direction.

Most of the work done up to now refers to the global diffusion in this system, i.e. the average diffusion over the whole phase space. However, more important are the local details of the diffusion. In particular, normal and anomalous diffusion coexist in many cases i.e. for the same value of $K$ of the mapping (\ref{stand}) certain initial conditions lead to normal diffusion, while other initial conditions lead to anomalous diffusion.

In the present paper we study the global and local diffusion for various ranges of the nonlinearity parameter $K$. \textbf{When} there exist islands of accelerator modes (\cite{chirikov1979universal}) the diffusion is ballistic globally (on the average). Furthermore, we find \textbf{some new cases of accelerator modes and we describe a method to derive higher order modes. Thus, we argue that the accelerator modes are much more common that the normal modes}.

 A problem considered in this study is the extent of the calculations in time. In some cases it is necessary to extend the calculations to extremely long times in order to find the correct values of the convergence of the diffusion exponent. E.g. we find that in cases where accelerator modes exist the mean diffusion of the whole phase space is always ballistic in the long run. On the other hand, the local diffusion of the chaotic regions is always normal (even if initial conditions are taken inside the extreme sticky regions), but in some cases it is manifested after a very large number of iterations, due to stickiness effects.  Ballistic diffusion appears locally only for initial conditions inside the accelerator mode islands, while outside the islands the diffusion is normal.   

In some references, the authors have claimed that even outside the accelerator mode islands, inside the stickiness region of the islands, there exists superdiffusion, i.e. nearby orbits are dragged by the accelerator mode islands (\cite{1982PhyD....4..425K},\cite{ishizaki1991anomalous},\cite{benkadda1997self},\cite{meiss2015thirty}). However, we find that this is true only for some time, and in general initial conditions in the chaotic region tend to normal diffusion after long enough time.  The anomalous phase of diffusion lasts longer in the close vicinity of the accelerator mode islands where stickiness effects are important.

Our paper is organized as follows. In section \rom{2} we calculate the global diffusion in various cases and we find that in general there are only two types of diffusion, normal and ballistic (anomalous) diffusion. For values of the nonlinearity parameter $K$, where accelerator mode islands exist (even if their size is extremely small compared with the whole phase space), the global diffusion of the whole phase space is dominated  by the ballistic motion of these islands. Then we find a correlation between the diffusion coefficient and the sizes of the accelerator mode islands of stability. In section \rom{3} we study in detail the local diffusion inside and outside normal and accelerator mode islands with emphasis on the sticky zones around these islands. In particular we examine cases where the normal and accelerator mode islands coexist and a case with maximum stickiness. In section \rom{4} we find analytically the characteristics of accelerator modes of period 2 and 4.  We present evidence
 that higher order accelerator modes appear in every interval $\Delta K$ although in the majority of the cases the diffusion is normal. Finally, in section \rom{5} we summarize our results.

\section{Calculation of global diffusion}

\subsection{Normal and Anomalous Diffusion}
The anomalous diffusion is related to the existence of accelerator
modes, i.e, orbits with initial conditions inside islands of stability
whose iterates extend to infinity if we ignore the modulo in $y$. The notion of "accelerator modes" was introduced by Chirikov \cite{chirikov1979universal}. More precisely, the images of an initial $y_0$ differ from $y_0$ by approximately an integer.  
These types of islands were studied in detail by  Contopoulos et al. \cite{contopoulos2005recurrence}
and they were named group II islands. They surround periodic orbits that are stable for values of $K$ belonging to intervals (\cite{chirikov1979universal}) :

\begin{equation}\label{accelst}
  (2\pi n)\leq K \leq \sqrt{(2\pi n)^2 +16}
\end{equation}
where $n$ is an integer.
For a little larger $K$ they generate higher order islands. In fact while the intervals (\ref{accelst}) are  about $\Delta K \approx 8/K$, the higher order intervals $\Delta K$ between successive bifurcations, of the accelerator mode islands, decrease by a factor $\approx$8.72 \cite{benettin1980universal}. Therefore the total extent of the interval $\Delta K$ that contain accelerator mode islands is:
\begin{equation}\label{8k}
\Delta K = (8/K)/(1-1/8.72)\approx 9/K
 \end{equation}

Normal diffusion in two dimensional mappings  is related to random
walk and in the limit, for very small steps, one obtains Brownian
motion (\cite{metzler2000random}). In this case, we can define the diffusion coefficient
which is a constant depending on the non-linearity parameter $K$
and is given by the relation:

\begin{eqnarray} \label{diff2}
D(K)= \frac{\langle(y-y_0)^2\rangle}{n}  (n\rightarrow \infty)
\end{eqnarray}
where $n$ is the number of iterations of the mapping (and it is a discrete time) and $\langle ~ \rangle$
denotes the average over a large ensemble of initial conditions
(e.g.  \cite{ishizaki1991anomalous}, \cite{manos2014survey})
\footnote{In some papers (e.g. \cite{afanasiev1990width},
\cite{afraimovich1997fractal}) a different definition of $D(K)$ is
given ($D(K)= \frac{\langle (x^2+y^2)\rangle}{n}$). However,
in general people consider the definition (\ref{diff2}), omitting
the ``modulo'' 1 with respect to $y$, but keeping the ``modulo 1''
with respect to $x$ in Eq. (\ref{stand})}. In general we need a large value of $n$ in order to 
ensure the convergence of the diffusion coefficient.

\begin{figure}
\centering
\includegraphics[scale=0.4]{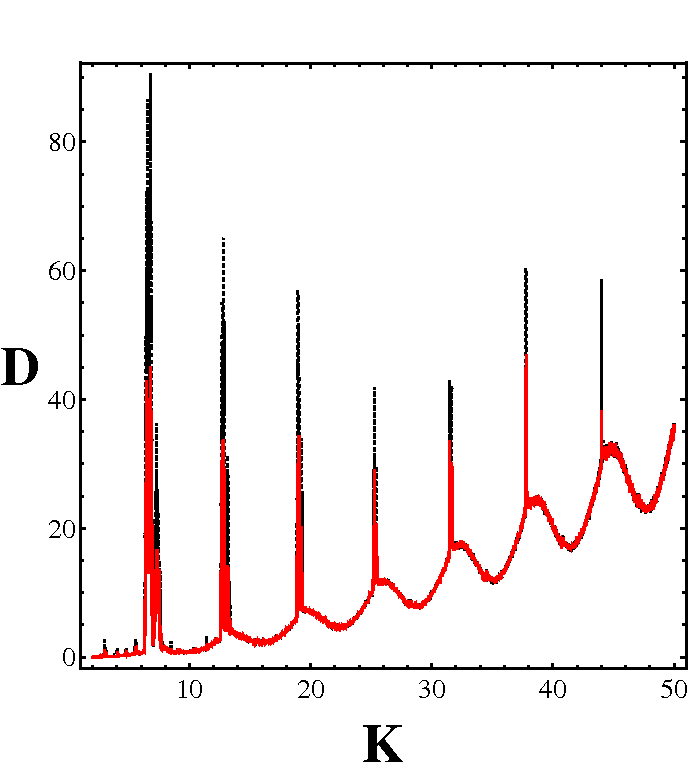}
\caption{ The diffusion coefficient $D(K)$ \textbf{derived} from Eq.(\ref{diff2})
as a function of the non-linearity parameter  $K$ (\textbf{gray (red in online version) solid curve)} for  $n=5\times 10^3$
iterations and black \textbf{dotted} curve for
$n=10^4$ iterations). The presence of accelerator modes in the intervals
defined by the Eq.(\ref{accelst}) generate anomalous diffusion
where the diffusion coefficient is no longer constant but it goes
to infinity with the number of iterations.} \label{difk}
\end{figure}

In Fig. \ref{difk} we plot the  diffusion coefficient $D$ calculated
by the relation (\ref{diff2}) as a function of the non-linearity
parameter $K$.   The number of iterations used for the red \textbf{solid} curve
is $n=5\times 10^3$.  On the other hand the black \textbf{dotted} curve
corresponds to  $n= 10^4$. We have used a grid of $10^4$ initial
conditions uniformly distributed on the entire initial phase space $[0,1]
\times [0,1]$ in order to calculate the mean value $\langle(y-y_0)^2\rangle$
for the specific number of iterations. The presence of accelerator
modes in some intervals defined by Eqs.(\ref{accelst}-\ref{8k}) generates
anomalous diffusion and the diffusion coefficient $D(K)$ is no longer constant but goes to infinity
when the number of iterations goes to infinity (\cite{ishizaki1991anomalous}). That is why the peaks of the black curve of Fig. \ref{difk}
have greater values than the peaks of the red curve. Similar figures can be found in 
\cite{zaslavsky1997self}, \cite{venegeroles2008calculation}, \cite{manos2014survey}.

\begin{figure*}
\centering
\includegraphics[scale=0.6]{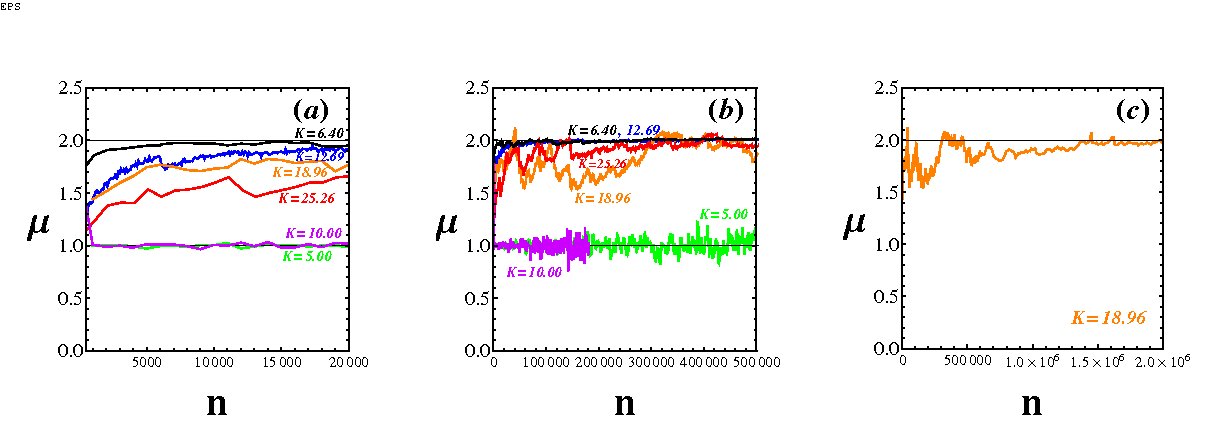}
\caption{(a) The diffusion exponent $\mu$ as a function of the
number of iterations $n$ for several values of the non linearity
parameter $K$ for a number of iterations up to $2 \times 10^4$. The
values of $\mu$ converge fast to $\mu=1$ in the case of normal
diffusion e.g. for $K=5.0$ and $K=10.0$. On the other hand in the
cases of anomalous diffusion (accelerator modes) the values of
$\mu$ converge to $\mu=2$ after a time of the order of $n=2 \times 10^4$
for $K=6.4$ and $K=12.69$, while the other cases of accelerator modes have not yet converged (b) Same as in (a) for a longer
time up to $5 \times 10^5$. We observe that the values of $\mu$ seem to converge around
the value of 2 after $n\approx 3 \times 10^5$ for all the values of $K$ related
to the accelerating modes while they converge to 
1 for values of $K$ related to the normal mode. (c) The case of $K=18.96$ calculated  for even larger values of $n$, where it is obvious that $\mu$  finally converges to the value 2.} \label{difm}
\end{figure*}

The generalized diffusion process for the standard map  is defined as follows:
\begin{eqnarray} \label{diff3}
\langle(y-y_0)^2\rangle=D_{eff} (K) n^\mu
\end{eqnarray}
where $D_{eff}$ is the effective diffusion coefficient and  $\mu$ is the diffusion exponent (or transport exponent) as it was introduced in \cite{zaslavsky2002chaos} and its convergence can be found for $n \rightarrow \infty$. In the case of normal
diffusion, where no accelerator mode islands exist, $\mu=1$ (\cite{ishizaki1991anomalous}, \cite{zaslavsky1997self}, \cite{benkadda1997self}, \cite{manos2014survey})  and the
diffusion coefficient $D(K)$ is constant and depends only on $K$. The theoretical value of $D(K)$ for the standard map in the case of normal diffusion is given in a number of papers (e.g. \cite{rechester1980calculation}, \cite{israilev1990simple}, \cite{ma1992lieberman}, \cite{zaslavsky1997self}, \cite{venegeroles2008calculation}, \cite{manos2014survey}) and it was tested and compared with the numerical results like the ones of Fig. (\ref{difk}). $D(K)$ increases as $K$ increases in general, with quasi-periodic variations, in the case of normal diffusion, but for a given $K$ it remains roughly constant as $n$ increases.

In the case of anomalous diffusion we have subdiffusion when
$0<\mu<1$ or superdiffusion if $1<\mu\leq 2$. The case where
$\mu=2$ is called ballistic transport, and it is associated with
the presence of accelerator modes. 

In Fig. (\ref{difm}) we have calculated the diffusion exponent
$\mu$ from the logarithmic slope of the average variances $\langle (y-y_0)^2\rangle$ as
a function of the number of iterations $n$, for various values of
$K$. The mean value of $\langle(y-y_0)^2\rangle$ is derived from a large
number of initial conditions ($5\times 10^4$) that cover the whole initial
phase space ($0<x<1$, $0<y<1$). Manos and Robnik in \cite{manos2014survey} derived a relation between
the values of $\mu$ corresponding to accelerator modes and the
non-linearity parameter $K$ (see Fig. 4 of Manos and Robnik in \cite{manos2014survey}),
after $n=5 \times 10^3$ iterations. But they claimed that the diffusion
exponent has well converged after $n=5 \times 10^3$ iterations, to various
values between $\mu=1$ and $\mu=2$, depending on the nonlinearity
parameter $K$. However we find  that for values of $K$ where
accelerator modes exist, the values of $\mu$ increase in general for $n > 5 \times 10^3$ (Fig.2a) and converge to $\mu=2$,
after a much larger number of iterations (Fig. \ref{difm}b). For
even larger values of $n$ the values of $\mu$  may undergo some small
fluctuations near the value $\mu=2$, but in general they converge to $\mu=2$ (Fig. \ref{difm}c).  On the other hand the values of
$\mu$ converge around the value of $1$ for values of $K$
corresponding to normal modes (normal diffusion) e.g. \textbf{for} $K=5$ and $K=10$ (Figs. \ref{difm}a,b). The
values of $K$ selected for the accelerator modes
($K=6.4,12.69,18.96,25.26$) correspond to the maximum peaks of
Fig.(\ref{difk}). The values of $\mu$ in Fig. \ref{difm} are global values that refer to the whole phase space. 

An explanation why the global value of the diffusion exponent $\mu$ for the whole phase space must always converge to the value of $\mu=2$, for values of $K$ where accelerator modes islands exist, is the following:\\
The $m^{th}$ image of the initial point $(x_0,y_0)$ of an accelerator periodic orbit of multiplicity $m$, is $(x_0(mod1), y_0+ml)$, where $l$ is an integer and $|l|\geqslant 1$. Then its $mn^{th}$ image is   $(x_0(mod1), y_0+nml)$ and $\langle(y-y_0)^2\rangle=m^2 n^2 l^2$, i.e. $\mu=2$ (from eq. \ref{diff3}). For initial conditions inside an accelerator mode island of stability, the value of $\langle(y-y_0)^2\rangle$ has small deviations from $m^2 n^2 l^2$ and thus again $\mu=2$. 

If there exist accelerator mode islands of stability in the phase space (for a certain value of $K$), even if their measure is extremely small, the global value of $\mu$ for the whole phase space, is again $\mu=2$. E.g.
if we assume that the 99$\%$ of the total area corresponds to normal diffusion with $D=\langle(y-y_0)^2\rangle/n$=constant, while only 1$\%$ of the total area corresponds to accelerator diffusion with $D_{eff}=\langle(y-y_0)^2\rangle/n^2$=constant, then the average rate of the diffusion coefficient $D_{all}$ for the whole phase space will be: 

\begin{equation}\label{approx}
D_{all}=\frac{\langle(y-y_0)^2\rangle}{n}\approx 0.99D +0.01 D_{eff}.n 
\end{equation}
and $D_{all}\rightarrow\infty$ as $n\rightarrow\infty$. In fact the $1/100$ of the phase space corresponding to  accelerator mode initial conditions dominates the total number of initial conditions when $n>100$.

In the above calculations we assume that the number of test particles (initial conditions) is large enough to populate sufficiently the acceleration region. However, in some cases the accelerator mode islands are so small that one needs to take a very large number of initial conditions in the whole phase space, in order to populate them. E.g. in a case considered in \textbf{the Appendix} the area of the islands is only of the order of $10^{-6}$ of the total phase space. Then, if we have, $10^4$ initial conditions in the whole phase space area, the probability of having one initial condition inside the island is only $1/100$.
Thus in order to populate smoothly the whole phase space we should use at least $10^8$ initial conditions which requires a very long computation \textbf{time}.

A way to avoid this difficulty is by taking a reasonable large number of initial conditions, for example $10^4$, but populate more densely the area of acceleration to clearly demonstrate its effect.

 In our numerical calculations, we \textbf{verify} that the mean value of $\mu$ for the whole phase space converges always to the value $\mu$=2, when there exist accelerator mode islands, after a large enough number of iterations.

\subsection{Correlation between the diffusion coefficient and the area of the islands}
In a previous paper \cite{contopoulos2005recurrence} we have calculated
numerically the area of the islands of stability in the standard
map as a function of the non-linearity parameter  $K$ (see Fig.8
of \cite{contopoulos2005recurrence}). In that paper we have named "group I" and "group III"
the islands of stability that are related to normal modes
(and normal diffusion) and "group II" the islands of stability
that are related to accelerator modes (and anomalous
diffusion). The shapes of those figures presented a fractal form
with infinite peaks and minima. The minima are due to the abrupt reduction of the areas  inside the last KAM
curves surrounding the islands of stability with increasing $K$, whenever a resonance is crossed.

\begin{figure}
\centering
\includegraphics[scale=0.5]{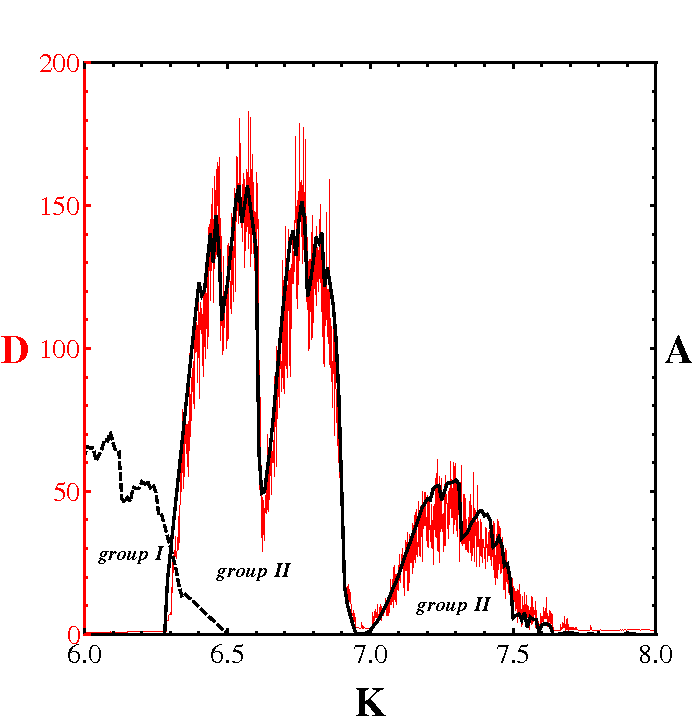}
\caption{The diffusion coefficient
$D(K)=\langle(y-y_0)^2\rangle/n$ (\textbf{gray (red in online version) curve})) calculated for $n=10^4$ iterations
 superimposed with the relative area $A$ of the
stability islands corresponding to the accelerator modes (group II in \cite{contopoulos2005recurrence}) for
values of $K$ in the interval $6.3\leq K \leq 7.7$
multiplied by a factor 2x$10^4$ (black curve).  The dashed black curve on the left refers to normal mode islands (group I in \cite{contopoulos2005recurrence}). For an interval of $K$ ($6.3\leq K \leq 6.5$) a coexistance of normal and accelerator mode islands takes place. } \label{difsur}
\end{figure}

 In Fig.\ref{difsur} the relative areas of the islands of stability are shown (black curve), for $6<K<8$, together with the diffusion coefficient $D$ calculated for $n=2\times 10^4$ iterations (red curve). 
This figure shows that there is a correlation between the diffusion
coefficient $D$ and the relative area $A$ of the accelerator mode islands of stability (or "group II" in \cite{contopoulos2005recurrence}). In order to match the two curves we have to multiply the areas $A$ by a normalizing factor $2 \times 10^4$. We observe that they present the same fractal form as functions of $K$. 

Figure \ref{difsur} contains accelerator mode islands in the range of eq. (\ref{accelst}) with $n=1$. Beyond $K=[(2 \pi)^2 +16]^{1/2}\approx 7.4$ the periodic orbit of group II becomes unstable but generates by bifurcation a stable period 2 family. (For $n=1$ this is seen in Fig. 21 of 
\cite{contopoulos2005recurrence}). For a little larger $K$ this family also becomes unstable and generates a period-4 family. As $K$ increases further we have a cascade of infinite period doubling bifurcations of periodic orbits that are stable in intervals decreasing by a factor  of $\delta \approx 8.72$ at every bifurcation (\cite{benettin1980universal}). Then beyond a limiting $K_{lim}$ slightly larger than $K=7.7$ we have an infinity of unstable periodic orbits. These are still accelerator modes but their set is of measure zero and they do not trap any set of points of finite measure that should give anomalous diffusion. For $K$ larger than $K_{lim}$ the diffusion is normal until we reach values of $K$ satisfying eq. (\ref{accelst}) for $n=2$ where  new accelerator modes exist and so on.
\begin{figure}
\centering
\includegraphics[scale=0.25]{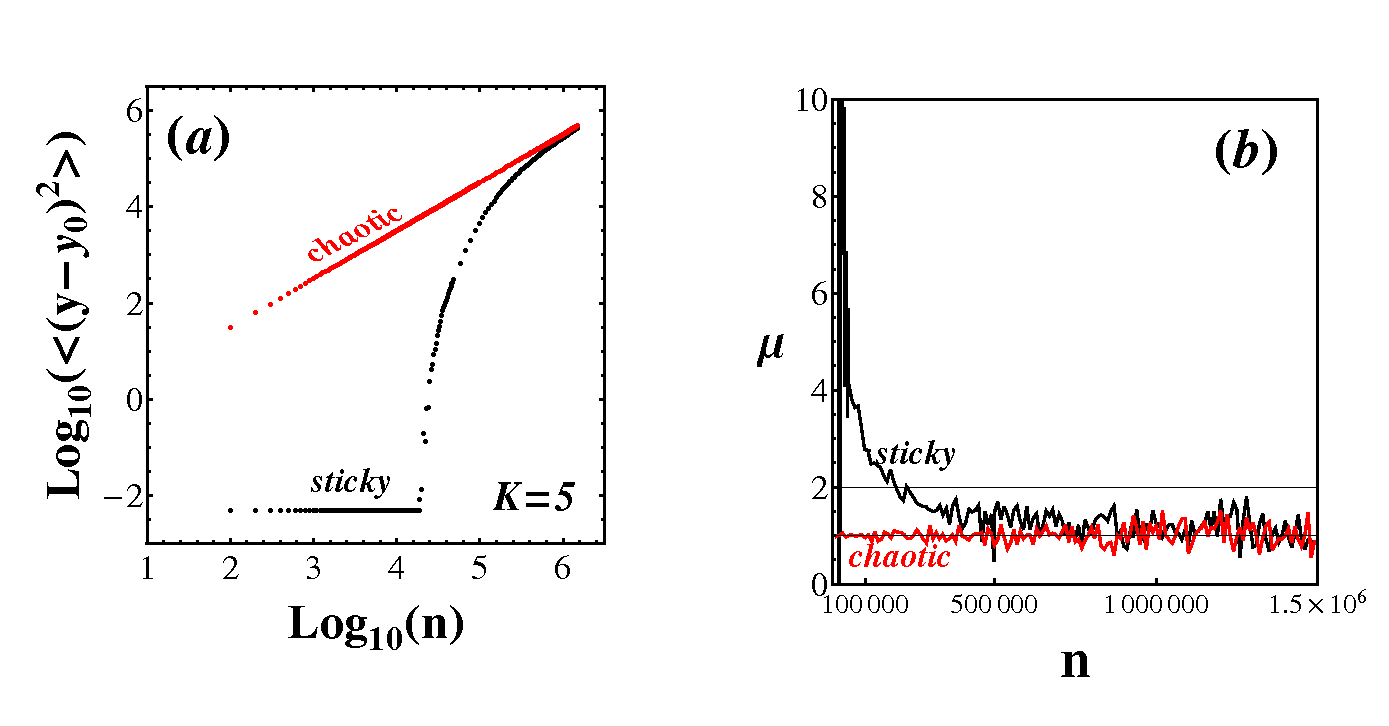}
\caption{ (a) The evolution of $\langle(y-y_0)^2\rangle$ as a function of the number of iterations $n$, in a logarithmic scale for a set of initial conditions inside the large chaotic sea (\textbf{gray (red in online version) curve}) and inside the sticky region of the normal mode island (black curve) for $K=5$. (b) The slope of the curves in (a) as a function of the number of iterations $n$, corresponding to the diffusion exponent $\mu$. The initial conditions inside the large chaotic sea converge fast to $\mu=1$, while the initial consitions inside the sticky region begin with $\mu=0$ during the time of stickiness and then, after a transient time of abrupt increase of the slope it converges to $\mu=1$ which corresponds again to normal diffusion.  } \label{stick}
\end{figure}
Another point of interest refers to the form of the curve $D(K)$ of Fig. \ref{difsur}. We notice that the curve seems to go to zero for $K\approx6.96$. This point refers to a resonance $1/3$ and it is well known that in this case the area of the island goes to zero (e.g. see Fig. 2.44 of \cite{contopoulos2013order}). Around this periodic orbit there
exists a small area with points that seem to present ballistic motion like the accelerator mode periodic orbit, but for a finite number of iterations (that depends on the distance from the periodic orbit, see subsection \rom{3}.B below) and then for a larger number of iterations they follow normal mode motion with diffusion exponent $\mu\rightarrow 1$ for $n\rightarrow\infty$.
We have tested this in the particular case of $K=6.96264956$ where the size of the accelerator mode island goes to zero (see Fig. \ref{difsur}). On the other hand, whenever there exists an accelerator mode island, even when it occupies an extremely small area compared to the whole phase space, the mean value of the diffusion exponent of the whole phase space will finally converge to 2, when the number of iterations $n$ goes to infinity. In order to find this, initial points that populate these small islands must be included in the calculations.

\section{Calculation of local diffusion}

\subsection{Diffusion in and around  normal mode islands}

The calculation of the diffusion coefficient as an average over the whole phase space does not give enough information about the diffusion process.  That is why in what follows we compute the diffusion exponent $\mu$ locally in specific regions of the phase space. For example, inside a normal mode island the diffusion coefficient tends to zero. In fact, the value of $\langle(y-y_0)^2\rangle$ is finite and constant \textbf{and therefore from eq. (\ref{diff3}), $\mu \rightarrow 0$, when $n$ tends to infinity}.

If we calculate the diffusion exponent $\mu$ from individual orbits we find a large scatter of values. Only if we take a large number of nearby orbits in a region of very small size  we can find reliable local values of $\mu$.   In our calculations we use $10^{4}$ initial conditions inside a small region of the order of $10^{-5}$ or $10^{-6}$. Using a larger number of initial conditions we have less fluctuations of the computed values. However, increasing a lot the number of initial conditions can be very time consuming and we do not gain more information. 

 The time evolution of the diffusion exponent $\mu$ for initial conditions inside a small area in the chaotic sea (red curve) and the in the sticky region around an island of stability that corresponds to normal diffusion (black curve) for $K=5$, is shown in Fig. \ref{stick}.
In Fig. \ref{stick}a we see the evolution of $\langle(y-y_0)^2\rangle$ as a function of the number of iterations $n$, in a logarithmic scale, for the two cases while in Fig. \ref{stick}b  
the diffusion exponent $\mu$ (which is the slope of the curves of Fig. \ref{stick}a) is plotted as a function of the number of iterations for the two cases (red curve corresponds to the chaotic region and black curve to the sticky region of the normal mode island). During the stickiness time $\mu$ has a mean value around zero, following the same behaviour with the initial conditions inside the normal mode island \textbf{(black curve)}. Then during the transition from the sticky zone to the large chaotic sea an abrupt rise of $\mu$ takes place  and for larger $n$ the value of $\mu$ presents a slow decrease on the average and finally it converges towards the value of 1. On the other hand, for the case of initial conditions in the large chaotic sea,  $\mu$ converges fast to the value of 1 \textbf{(red curve)}.   

One may argue that if an initial condition is in the sticky region of an island of stability (and therefore its initial values of the diffusion exponent $\mu$ are close to zero), then after a Poincar$\acute{e}$ recurrence time its images will come again close to the island and its values of $\mu$ will be again close to zero. However, different initial conditions (no matter how close to each other they are) have different Poincar$\acute{e}$ recurrence times (see discussion of this in the next subsection). Therefore, if we take a large number of initial conditions, their average long run diffusion coefficients will not come back close to zero, but they will remain close to $\mu=1$.

 The mean value of $\mu$ for the whole phase space in the normal diffusion cases $K=5$ and $K=10$ has well converged to $\mu=1$ even for a small number of iterations $n$ (see Fig. \ref{difm}b).

\subsection{Diffusion in and around accelerator mode islands }

\begin{figure}
\centering
\includegraphics[scale=0.25]{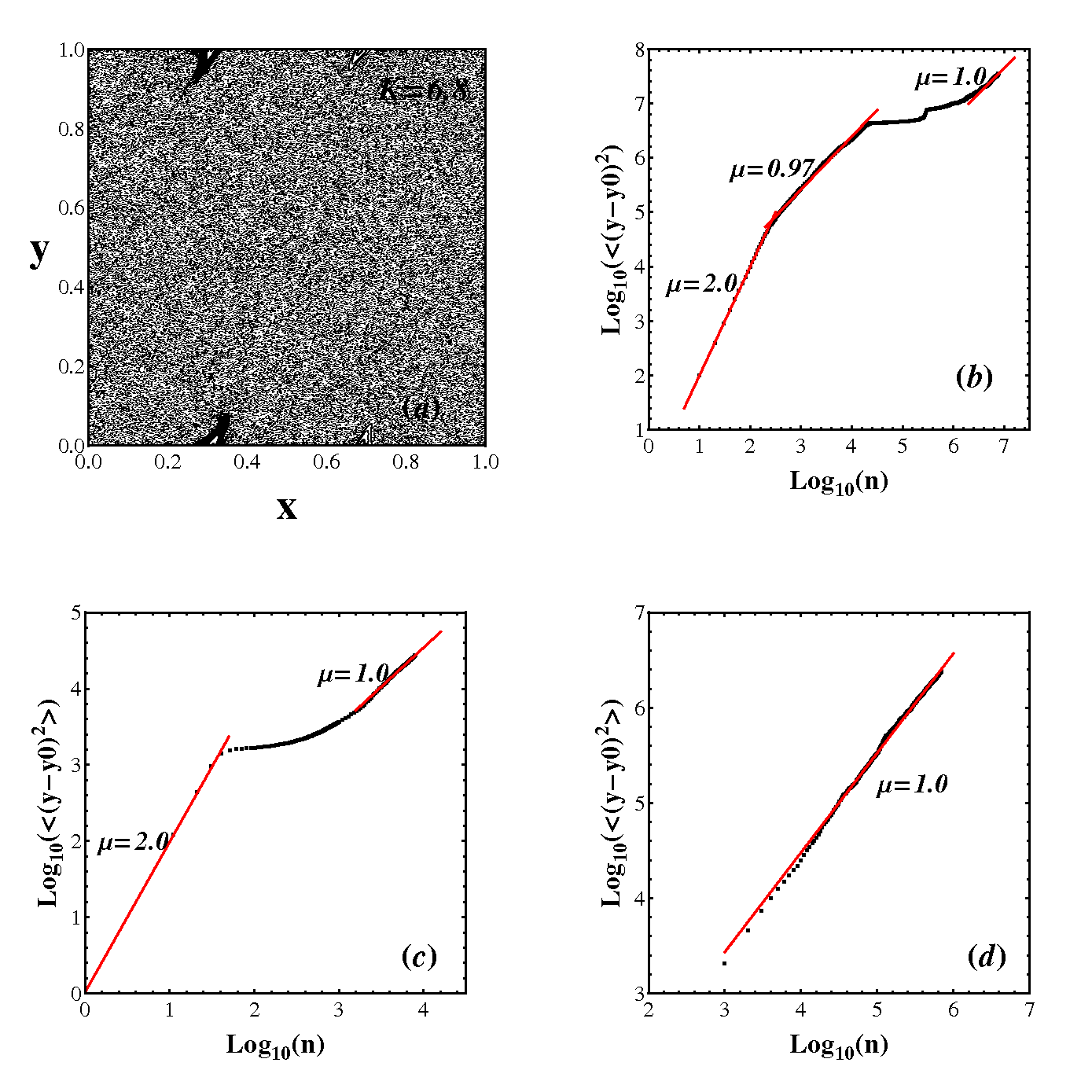}
\caption{(a) The phase space for $K=6.8$ with accelerator mode islands (calculations are made for the sticky, black region, around the islands of stability) (b) The evolution of $\langle(y-y_0)^2\rangle$ as a function of the number of iterations $n$,  in a logarithmic scale for a set of initial conditions inside the extreme sticky region of (a). The diffusion is anomalous (ballistic) for the time interval of stickiness around the accelerator mode island  ($n_{st}\approx 10^3$) where the diffusion exponent is equal to 2 ($\mu=2$), but after a transient time it becomes normal with diffusion exponent $\mu=1$  (c) Same as in (b) but for a set of initial conditions in the outer sticky region. Here the ballistic motion lasts for a shorter time ($n_{st}\approx 10^2$) (d) Same as in (b) for a set of initial conditions far from the sticky region and in the large chaotic sea. Here the diffusion is mormal almost from the beginning of time.} \label{k68}
\end{figure}

\begin{figure}
\centering
\includegraphics[scale=0.35]{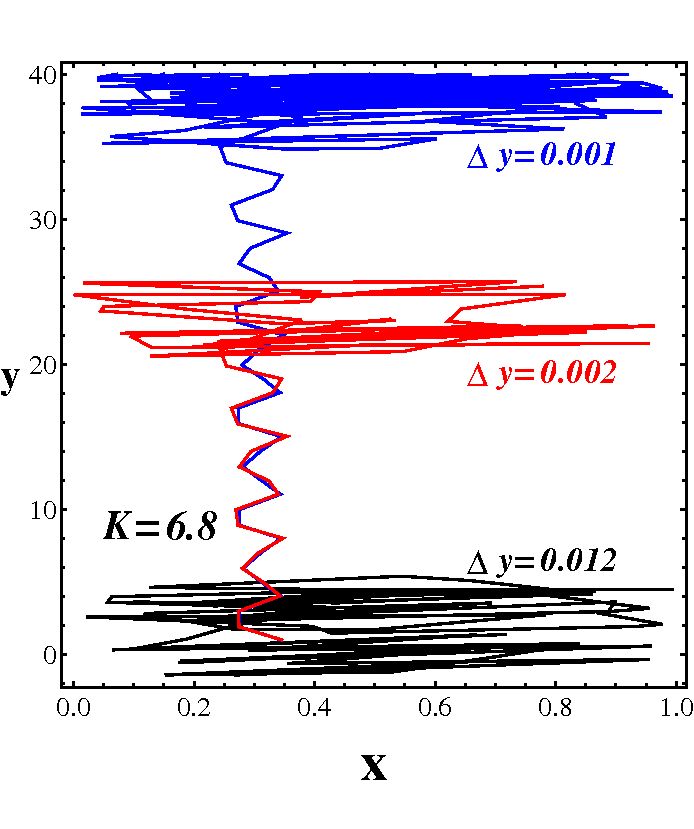}
\caption{ The displacements of $y$ for 3 initial conditions for $K=6.8$. The black curve corresponds to an initial condition outside the sticky zone of Fig. \ref{k68}a with distance $\Delta y=0.012$ from the last KAM curve of the island of stability which diffuses normally, while the other initial conditions (red and blue \textbf{in online version}) which are closer to the sticky zone of Fig. \ref{k68}a, (with distance $\Delta y=0.002$ and $\Delta y=0.001$ respectively) exhibit ballistic motion for a while, dragged by the accelerator mode island and then continue with normal diffusion.}
 \label{stick68}
\end{figure}

\begin{figure*}
\centering
\includegraphics[scale=0.25]{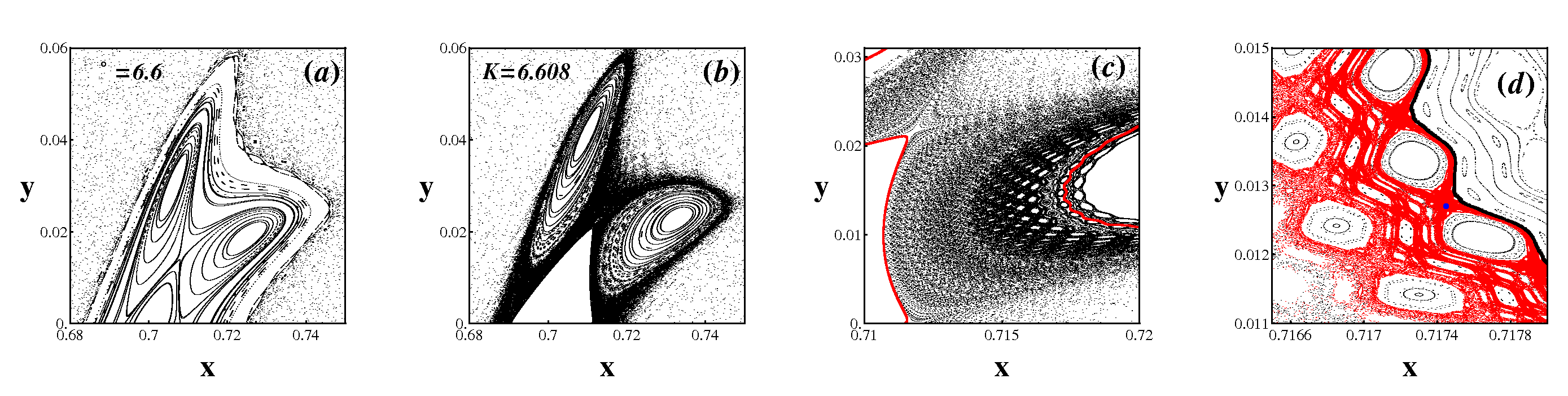}
\caption{(a) A part of an accelerator mode island for $K=6.6$ around a periodic orbit with $x\approx0.7,~y=0$. This island is surrounded by four accelerator mode islands surrounding a period-4 periodic orbit (only two of the four islands are shown here) that has bifurcated from the central orbit. These orbits lie inside the last KAM curve and do not communicate with the large chaotic sea (b) The same as in (a) for $K=6.608$. Here the period-4 periodic orbits  are inside the large chaotic sea and extreme sticky regions surround them. The area inside the last KAM curve has decreased abruptly (c) A zoom of the sticky region around the accelerator mode islands outside the last KAM curves (\textbf{thick gray (red in online version) curves}) of the islands of stability. (d) The extreme sticky region (\textbf{gray (red in the online version)}) outside the last KAM curve (thick black curve) of one of the four islands.  } \label{accel1}
\end{figure*}

\begin{figure*}
\centering
\includegraphics[scale=0.5]{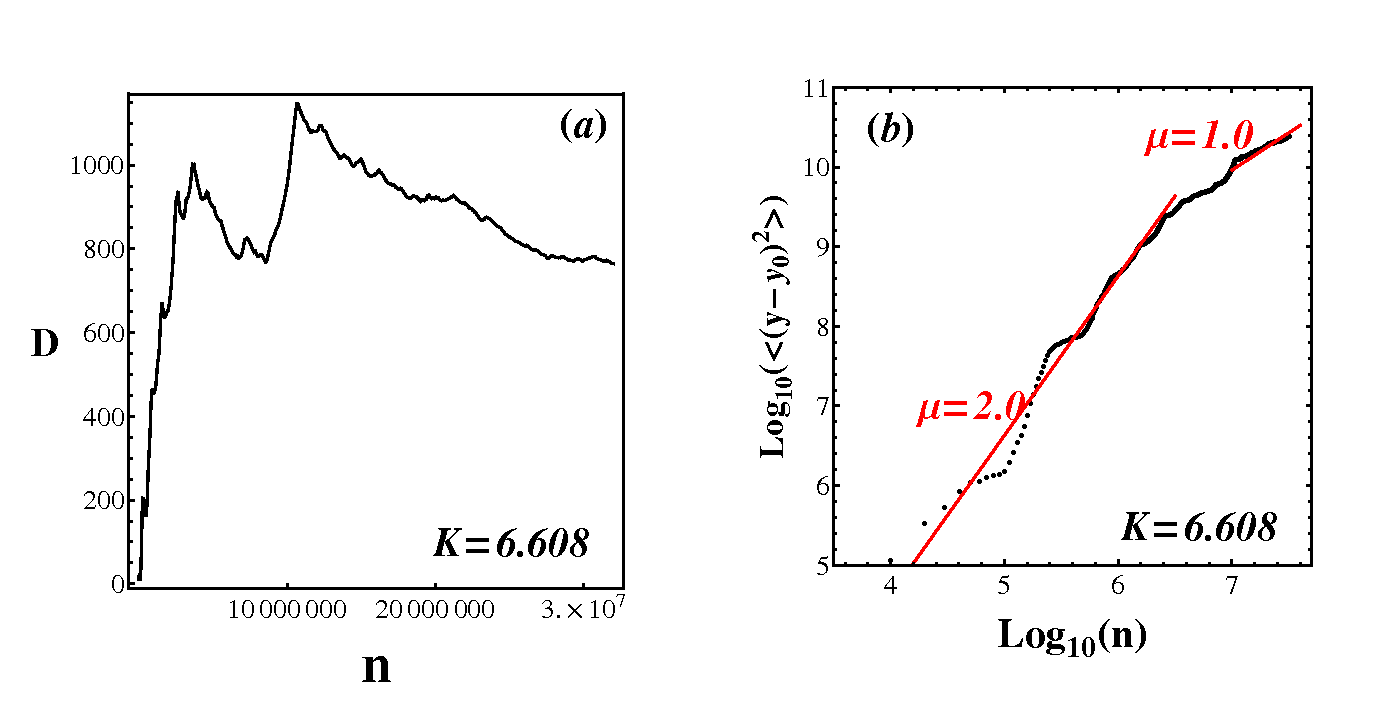}
\caption{(a) The diffusion coefficient $D$ as a function of time for the case of $K=6.608$ and a set of initial conditions inside the large chaotic sea far away from the sticky region of the acceleratot mode islands. The value of $D$ tends towards a constant value after more than 3x$10^7$ iterations. (b) The evolution of $\langle(y-y_0)^2\rangle$ as a function of the number of iterations $n$, in a logarithmic scale. The slope of this curve corresponds to the diffusion exponent $\mu$, which has a mean value of $\approx 2.0$ until $5 \times 10^6$ iterations while it converges to the value $\mu=1$, corresponding to normal motion, after 3x$10^7$ iterations.} \label{slope6608} 
\end{figure*}

\begin{figure}
\centering
\includegraphics[scale=0.4]{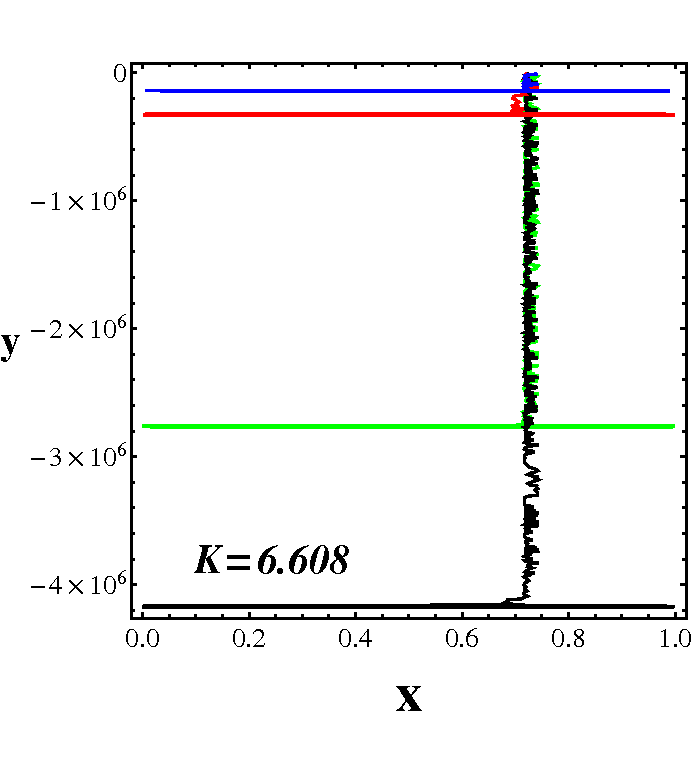}
\caption{The displacements of $y$ for several initial conditions of the phase space for $K=6.608$. The dragging of the accelerator mode island can last for very long times (the black curve corresponds to ballistic motion for a time of 4 x $10^6$ iterations, during which the displacement of $y$ is proportional to the number of iterations $n$), depending on how close the initial condition is to the island, but all initial conditions outside the accelerator mode island finally exhibit normal diffusion. }
 \label{stick6608}
\end{figure}

The diffusion exponent $\mu$ for initial conditions inside an accelerator mode island converges fast to the value of $\mu=2$. 

On the other hand, in the case of local initial conditions inside the sticky region around the accelerator mode islands, the diffusion exponent is  $\mu$=2 during a transient period of time (which is approximately equal to the time of  stickiness). During that time the chaotic orbits are dragged by the accelerator mode islands in a ballistic motion, but finally the orbit ends up in normal diffusion with $\mu=1$ for a large enough number of iterations. This is shown in Fig. \ref{k68} where accelerator mode islands exist for $K=6.8$. 

In Fig. \ref{k68}a, the whole phase space of the map (\ref{stand}) is shown using modulo 1 for both components $x$ and $y$ for $K=6.8$. The extreme sticky region around one of the two accelerator mode islands is shown (black region). Taking $10^4$  initial conditions in a very small box of size ($10^{-5}$x$10^{-5}$) we derived the local values of the diffusion exponent for long enough time, in order to ensure the convergence. In Fig. \ref{k68}b the evolution of the diffusion exponent is calculated from the logarithmic slope of the time evolution of  $\langle(y-y_0)^2\rangle$, for initial conditions in the extreme sticky region of the accelerator mode island (black regions in Fig. \ref{k68}a) at a distance $\Delta y=10^{-4}$ from the last KAM curve (\textbf{x=0.3}). During the time of stickiness  the motion is ballistic with $\mu=2$ (up to $n \approx 10^3$), but then after a transient period of about $10^6$ iterations, it finally follows a normal diffusion with $\mu=1$. The same is true for initial conditions in the outer sticky region (Fig. \ref{k68}c) and a distance $\Delta y=5 \times10^{-4}$ (\textbf{x=0.3}), where the anomalous diffusion with $\mu=2$ lasts for less time (up to $n \approx 10^2$) and the stage of normal diffusion starts earlier,
 after about $10^3$ iterations. Finally in Fig. \ref{k68}d where the initial conditions were taken far away from the sticky region and in the large chaotic sea (at a distance $\Delta y=0.4$ from the last KAM curve, \textbf{x=0.3}), the diffusion is normal ($\mu=1$) from the beginning of our calculations.

 The displacements of $y$ are shown for different initial conditions in Fig. \ref{stick68}. The black (lower) curve corresponds to initial condition far away from the sticky region (with $\Delta y=0.012$ from the last KAM curve) and the diffusion is normal from the beginning, while the red and blue (upper) curves correspond to initial conditions that are closer to the sticky region  (with $\Delta y=0.002$ and $\Delta y=0.001$ from the last KAM curve respectively) and exhibit ballistic motion for a short transient time period.  Therefore, we conclude that for all initial conditions in the chaotic region, outside the accelerator mode islands of stability,  the diffusion will always end up being normal (with diffusion exponent $\mu=1$), after a transient time of ballistic motion that depends on the distance from the island of stability. 
 
\subsection{A case of maximum stickiness }

A particular case with a maximum stickiness around an accelerator mode island appears for $K_{max}=6.608$. Close to this value of $K$, the relative area of the stability decreases abruptly (see Fig. \ref{difsur}). In  Fig. \ref{accel1}a a small region of  the phase space for $K=6.6$ is plotted with modulo 1 for both $x$ and $y$. In this case there exists an accelerator mode island (near the axes $y$=0 and $y$=1) around the periodic orbit ($x\approx 0.7, y=0$ and  $y=1$)(there are two parts of the island due to modulo 1). 
Around this central island there are 4 secondary islands that have bifurcated from the central periodic orbit at $K\approx6.5938$. For this value of $K$ the area of the islands of stability is close to a maximum (see Fig. \ref{difsur}) and the 4-islands of stability are inside the last KAM curve and do not communicate with the large chaotic sea.  In  Fig. \ref{accel1}b the same area of the phase space is plotted for $K=6.608$, where the last KAM curve around the 4 islands has been destroyed and these islands have entered the large chaotic sea. The area inside the new last KAM curve has decreased abruptly. A large sticky region is generated now around the period-4 islands of stability. The sticky region around the 4-islands in Fig. \ref{accel1}a  is quite small in comparison with the sticky region around the 4-islands in Fig. \ref{accel1}b.
 The sticky regions in a greater focus are shown in Fig. \ref{accel1}c together with the last KAM curves (red curves) that separate the islands from the sticky chaotic region.  Finally in Fig. \ref{accel1}d  we plot in red the extreme sticky region around one of the 4 islands, where the last KAM curve is shown now in black. 

If we take an initial condition in the chaotic sea, far away from the sticky zones around the islands of stability and plot the time evolution of the diffusion coefficient $D$ we see (Fig. \ref{slope6608}a) that for a very long time  (about 3x$10^{7}$ in this case) the effective diffusion coefficient $D$  varies considerably. But after this transient period, the diffusion coefficient $D$ reaches an approximate constant value. In Fig. \ref{slope6608}b we plot the logarithm of the average value of the displacement $\langle (y-y_0)^2\rangle$ as a function of the logarithm of the number of iterations $n$ for a set of $10^4$ initial conditions. The slope of this curve corresponds to the diffusion exponent $\mu$. The diffusion exponent $\mu$ has a mean value equal to 2 for a time period of $\approx 5 \times 10^6$ iterations, which means that during this period of time the chaotic orbit is dragged by the accelerator mode islands. However, after about 3x$10^7$ iterations the value of $D$ is stabilized, and the exponent $\mu$ converges to the value 1. 
 
In every case, the ballistic motions of initial conditions close enough to the accelerator mode islands last only for a transient time period that can be very long sometimes, but  they will always end up with normal motion and the diffusion exponent will converge to the value of 1. This statement can be confirmed by Fig. \ref{stick6608}, where we plot the  displacements of $y$  with time, for some initial conditions for $K=6.608$. 
The initial conditions that are close to the sticky region produce ballistic motion initially, i.e. $y$ is proportional to the time $n$ for a while, dragged by the accelerator mode island but then continue with normal diffusion with small variations of $y$. This transient "dragging time" of ballistic motion may last for very long time, depending on how close to the island of stability the initial conditions are. This is seen in Fig. \ref{stick6608} where the initial condition corresponding to the black curve exhibits ballistic motion for more than 4x$10^6$ iterations (for that time the $x$ component stays located close to $x$=0.7) but then it ends up with normal diffusion. 

Besides the "dragging time", there are three important times: (a) the escape time (or "initial stickiness time") (b) the Poincar$\acute{e}$ recurrence time, and (c) the Lyapunov time.
Below we discuss the relevance of these times to the diffusion problem. We consider orbits in the sticky domain while both x and y are taken modulo 1.

(a) The escape time (or "initial stickiness time") gives the time needed for an orbit to escape from the sticky domain around an island of stability to the large chaotic sea. This is directly related to the dragging time, because after the orbits escape into the large chaotic sea (and before returning \textbf{again} in the sticky region) they are not dragged any more by the accelerator islands. Thus, roughly speaking before the escape time the diffusion is anomalous ($\mu \approx$2) while after the escape time the diffusion will finally converge to normal ($\mu$=1). A detailed discussion of the escape/stickiness times in the standard map has been made in     
\cite{2008IJBC...18.2929C}, \cite{2010CeMDA.107...77C}. In \cite{2010IJBC...20.2005C} we have \textbf{found} that the stickiness time outside an island of stability increases exponentially as we approach the island. Furthermore, this increase becomes superexponential very close to the boundaries of the island of stability as it was proved by Giorgilli and Morbidelli in \cite{1995JSP....78.1607M}. Whenever the chaotic orbits  are found in this region of stickiness they remain there for a time that corresponds to the stickiness time of this region, which depends on the distance from the last KAM curve of the island.  

(b,c) The Poincar$\acute{e}$ recurrence time is also important because after such a time an orbit returns close to its initial position and therefore it follows a similar evolution. i.e. whenever it returns in the sticky region it is dragged again along the accelerator islands for about the same time (the stickiness time). However, nearby orbits deviate considerably at a rate that depends on the Lyapunov time, thus they have quite different individual Poincar$\acute{e}$ recurrence times. In fact, the distribution of a set of initial conditions, inside the sticky zone of an island of stability, changes a lot  after an average Poincar$\acute{e}$ reccurence time  (see Appendix C of \cite{2010IJBC...20.2005C}).  As a consequence the new dragging time intervals of a set of initial conditions appear at different times and the average value of $\langle (y-y_0)^2 \rangle $ of this set does not increase in an anomalous way. In fact, because of the dispersion due to the \textbf{positiveness of the} Lyapunov exponent the average density of the points tends to become constant (see Fig. 29 of \cite{2010IJBC...20.2005C} and Fig. 13 of \cite{1995NYASA.773..145C}) after a very long time, and there is no distinction any more, \textbf{in the distribution of particles} between the large chaotic sea and the sticky domain. As a consequence a representative ergodic orbit fills uniformly both the large chaotic sea and the sticky zone. Therefore, in the long run it stays on the average in the chaotic sea and in the sticky zone for times $t_{rec}$ and $t_{st}$ respectively, that are proportional to their areas, $A_{ch}$ and $A_{st}$. Thus roughly  $t_{st}/t_{rec} \sim$ $A_{st}/A_{ch}$. In most cases the ratio $A_{st}/A_{ch}$ is very small and is the ratio $t_{st}/t_{rec}$.

\begin{figure}
\centering
\includegraphics[scale=0.5]{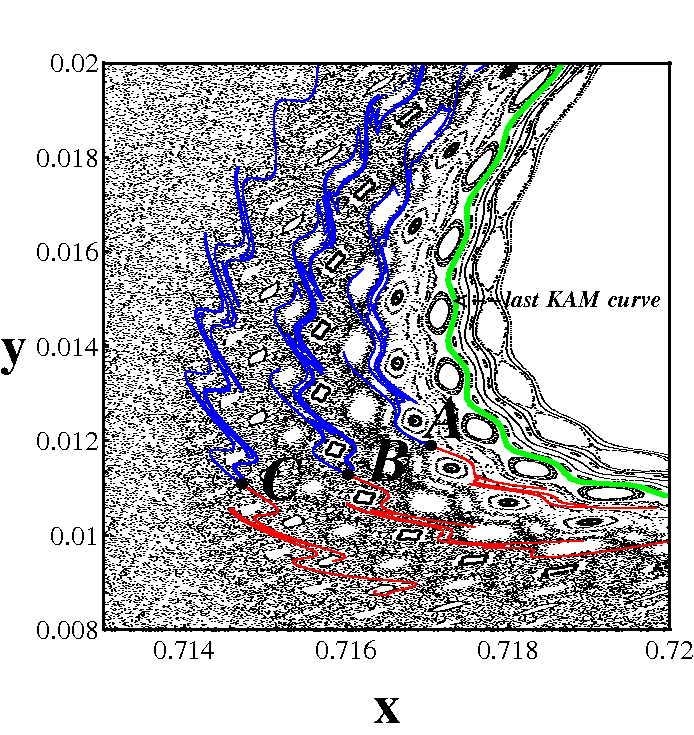}
\caption{ Three different unstable periodic orbits of multiplicity 23 (point A), 25 (point B) and 28 (point C) with their corresponding unstable manifolds (\textbf{ dark and light gray (blue and red in the online version) curves}) in the extreme sticky region of the accelerator mode island for $K$=6.608. The light gray (\textbf{green in the online version}) thick curve is the last KAM curve surrounding the island of stability. Initial conditions on these manifolds have different times of escape from the sticky zone and different times of Poincar$\acute{e}$ recurrence time (see text). The Poincar$\acute{e}$ recurrence times are in general much longer than the corresponding stickiness times.  }
 \label{man6608}
\end{figure}

In Fig. \ref{man6608} three different unstable periodic orbits of multiplicity 23 (point A), 25 (point B) and 28 (point C) are plotted together with their corresponding unstable manifolds (blue and red curves) inside the deep sticky regions of Fig. \ref{accel1}c. All initial conditions inside the sticky region will have to follow nearly parallel paths to the unstable manifolds of the unstable periodic orbits in order to reach the large chaotic sea. This is the reason of stickiness (see \cite{2010CeMDA.107...77C}), as the unstable manifolds of the unstable periodic orbits present many oscillations inside the sticky zone (around the islands of stability) before entering into the large chaotic sea. Initial conditions taken in a distance $ds=10^{-6}$ from these points (A,B and C), along the unstable asymptotic curves, have different escape times (or initial stickiness times) and different Poincar$\acute{e}$ recurrence times, as mentioned above. An initial condition  in a distance $10^{-6}$ from the point A has escape time $t_{esc}(A)=5 \times 10^5 $, while the Poincar$\acute{e}$ recurrence time in the vicinity of this point (inside a box of dimension $10^{-3}$) is $t_{rec}(A)=4 \times 10^9$ iterations.  The same values for the point B are :$t_{esc}(B)=1.8 \times 10^5 $ and $t_{rec}(B)=5.8 \times 10^8$ and for the point C are : $t_{esc}(C)=1.2 \times 10^5 $ and $t_{rec}(C)=3.1 \times 10^7$. These are representative values for these three different regions of the sticky zone and become larger for initial conditions closer to the last KAM curve of the island of stability. As a consequence, even if we take a large number of initial conditions inside a very small domain in the sticky region, the Poincar$\acute{e}$ recurrence times will be similar but not the same and although all the chaotic orbits with initial conditions inside this small region will stick for similar time intervals around the islands of stability, they will not arrive   there simultaneously. Therefore, the chaotic orbits inside the whole sticky region are not dragged in a ballistic motion at the same time. On the other hand, the  Poincar$\acute{e}$ recurrence time is always much larger (some orders of magnitude) than the stickiness time around the accelerator mode islands. This is true even in the regions very close to the last KAM curve, where the stickiness time becomes superexponential. This is why after a long enough time, a set or orbits with initial conditions inside a small domain of the sticky region is diffused normally and the corresponding mean value of the diffusion exponent converges to 1.

\subsection{Comparison with other studies}

It is of interest here to compare our results with previous papers of a number of authors.

Venegeroles in \cite{venegeroles2008calculation} has found theoretically that, under certain assumptions, the chaotic component of the standard map in the case where accelerator mode islands exist diffuses with an exponent $\mu\rightarrow 1.5$ due to trapping of trajectories in the vicinity of the islands and in \cite{2009PhRvL.102f4101V} he conjectures the universality of this value by presenting a number of other papers of 2-D mappings and Hamiltoninan systems with numerical evidence for values of $\mu$ close to 1.5 (references within \cite{2009PhRvL.102f4101V}). Moreover, Benkadda et al. in \cite{benkadda1997self} have found numerically, for the standard map, that in the sticky region of an accelerator mode island the diffusion exponent is $\mu \approx 1.44$, and in \cite{1998PhyD..116....8L} Leboeuf has found also that, for the case of the 2-D area preserving kicked Harper map, the average diffusion exponent $\mu$ for the chaotic part of the phase space (outside the islands) is close to the value 1.4. Finally, Meiss in \cite{meiss2015thirty} assumes that "an accelerator island will drag nearby chaotic orbits along, so that even when the ensemble average does not include the accelerator island, the effect is that $\mu >$1."
 However, in all these studies, the calculations were made for times of the order of $10^6$ iterations, while our calculations were made for more than $10^7$ iterations in order to find that the value of $\mu$ converges to a value equal to 1. For example, in Fig. \ref{slope6608} we see that for $n=10^6$ the value of $D$ and the corresponding diffusion exponent $\mu$ is very far from its "final" convergent state which appears beyond $n=3 \times 10^7$ (see also Fig. \ref{slk64} below).
 
\textbf{Another possibility is that $\langle (y-y_0)^2 \rangle \propto n ln(n) $, in the chaotic region of the phase space when accelerator modes exist, as it was found in a different problem by Jones and Young in} \cite{1994JFM...280..149J}. \textbf{In our cases we found that the relation $\langle (y-y_0)^2 \rangle \propto n $ is slightly better than $\langle (y-y_0)^2 \rangle \propto n ln(n) $  after a long time. However, a definite conclusion would require much longer calculations}.

A paper that makes very long time calculations ($10^{10}$) was published recently by Das and Gupte (\cite{2017PhRvE..96c2210D}) who studied the diffusion properties of a six-dimensional map. These authors gave the distribution of the diffusion exponents $\mu$ for different regimes of the phase space and found a clear boundary between the normal and superdiffusive regions. Namely they found two main peaks of the diffusion exponent $\mu$ at $\mu$=1 (normal diffusion) and $\mu$=2 (ballistic motion) and an insignificant contribution of other values in between, a result that strengthens our own results.

As regards the global diffusion exponent, Zaslavsky et al. in \cite{zaslavsky1997self}  have found, for the web map and a certain value of the nonlinearity parameter (where there exist accelerator mode islands) that the mean value of the diffusion exponent, for the whole phase space,  is $\mu$=1.26. 
 However, we have shown in subsections \rom{2}.A and \rom{3}.B that, when accelerator mode islands exist, the mean value of the diffusion exponent $\mu$ of the whole phase space is dominated by the regions inside these islands and so it converges to the value $\mu=2$. 
 
 Concluding, we may say that the diffusion exponent in the case of the standard map, is equal to 2 (ballistic motion) when accelerator mode islands exist, either locally (inside the accelerator mode islands) or globally for the whole phase space and equal to 0 locally (inside the normal islands of stability). On the other hand the diffusion exponent converges to 1 (normal diffusion) globally when no accelerator mode exists and locally for all chaotic regions outside the islands of stability, including the sticky regions (provided that one takes a large enough number of iterations).  

\subsection{Coexistence of normal and accelerator modes}

\begin{figure}
\centering
\includegraphics[scale=0.25]{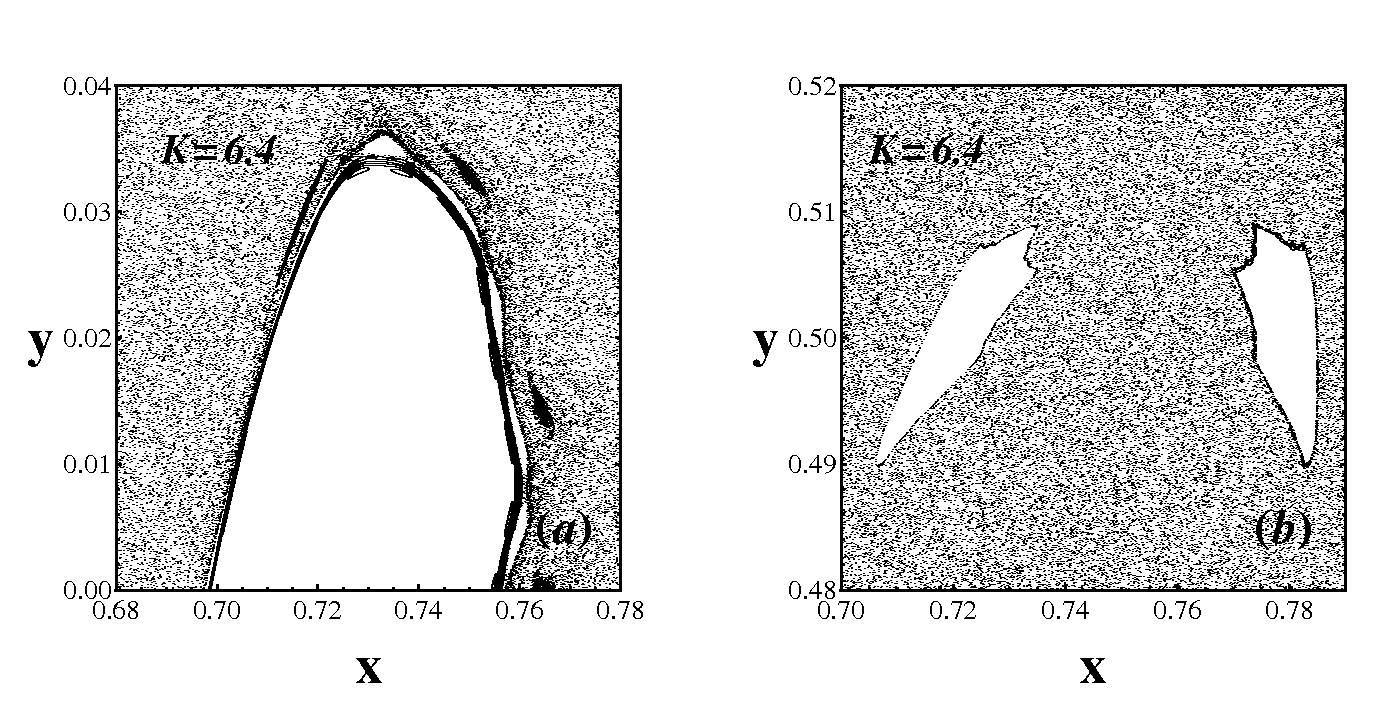}
\caption{(a) The accelerator mode island and the sticky region
around it and (b) Normal mode islands. The sticky region around one of them is shown. Both (a) and (b) are for the same nonlinearity parameter
$K=6.4$ where the normal mode and the accelerator mode islands coexist.}
 \label{noraccel}
\end{figure}

\begin{figure*}
\centering
\includegraphics[scale=0.5]{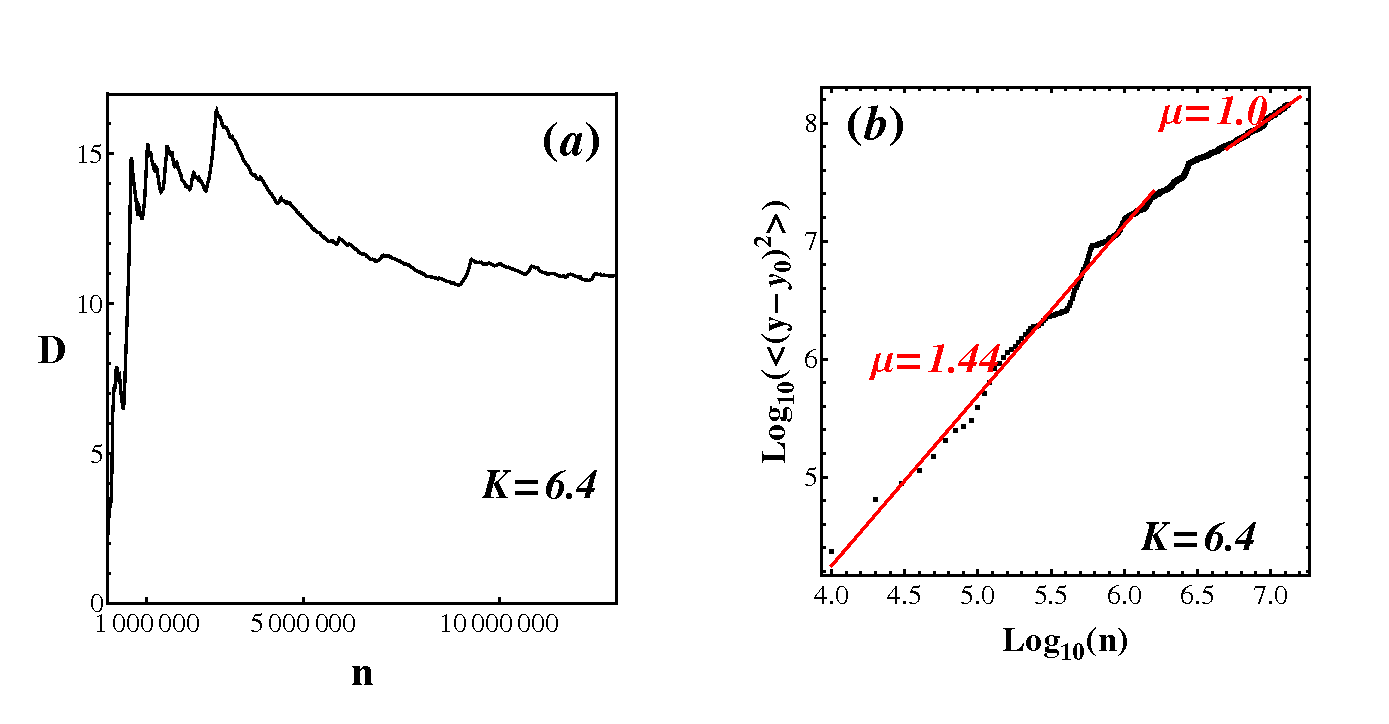}
\caption{(a) The diffusion coefficient $D$ as a function of time for the case of $K=6.4$ and a set of initial conditions inside the large chaotic sea far away from the normal or accelerator mode islands. The value of $D$ tends towards a constant value after more than $10^7$ iterations. (b) The evolution of $\langle (y-y_0)^2 \rangle$ as a function of the number of iterations $n$, in a logarithmic scale. The slope of this curve corresponds to the diffusion exponent $\mu$, which has a mean value of $\approx 1.44$ until $10^6$ iterations while it converges to the value $\mu=1$, corresponding to normal motion, after $10^7$ iterations.}
 \label{slk64}
\end{figure*}

An example of coexistence of normal and accelerator mode islands appears for $K=6.4$ (see Fig. \ref{difsur}). Figure \ref{noraccel}a shows the upper part of an acceleration mode island (one of the two period-2 islands) and Fig. \ref{noraccel}b shows two normal mode islands (each one belonging to a different period-2 island) for the same value of $K$. 
In fact these islands are formed around two period-2 periodic orbits that were generated after two bifurcations from a central periodic orbit ($x_0=0.5$, $y_0=0$) that exists for all $K>$0. 

The mean value of the global diffusion exponent $\mu$ in the case $K=6.4$ for the whole phase space converges to $\mu$=2, because of the existence of the accelerator mode islands (see Fig. \ref{difm}) but here we investigate the local diffusion exponent for initial conditions in the chaotic region, outside the islands of stability. 

 Chaotic orbits in the sticky zone around the accelerator mode island exhibit ballistic motion for a transient time period, as shown in the previous paragraph but the diffusion exponent $\mu$ will finally converge to the value $\mu=1$ that corresponds to normal diffusion, after some transient fluctuations between the values 1 and 2. The width of these fluctuations depends on the number of initial conditions taken and becomes smaller for larger numbers of initial conditions. On the other hand, chaotic orbits in the sticky region of the normal mode islands, have diffusion exponent close to $\mu$=0, as explained in subsection \rom{3}.A during their time of stickiness and then they follow the same behaviour as the initial conditions in the sticky region around the accelerator mode islands, i.e the diffusion exponent will again converge to the value 1.  
 
 In Fig. \ref{slk64}a the evolution of the diffusion coefficient $D$ is plotted      
as a function of time for $10^4$ initial conditions inside a small box of size ($10^{-5}$x$10^{-5}$) in the large chaotic sea,  away from the sticky zones of the normal and accelerator mode islands of stability.
 The value of $D$ has a very different behaviour in successive intervals of time before converging in a constant value after a relatively long time i.e. after approximately $10^7$ iterations. In  Fig. \ref{slk64}b
the evolution of 
 $\langle (y-y_0)^2 \rangle$ is plotted as a function of the number of iterations $n$, in a logarithmic scale. The slope of this curve corresponds to the diffusion exponent $\mu$. During an interval of time of about $10^6$ iterations from the beginning of our calculations the diffusion exponent has a mean value of $\mu$=1.44 (a value that has been given in many references as the final value of convergence of $\mu$, as pointed out in the previous paragraph). However, this value is transient and is due to the initial stickiness time when the chaotic orbit is dragged by the accelerator mode, while after about $10^7$ iterations the mean value of $\mu$ converges to the value of 1, manifesting the transition to the normal diffusion.

\section{Further accelerator modes}
\subsection{Accelerator modes of period 2}

\begin{figure*}
\centering
\includegraphics[scale=0.4]{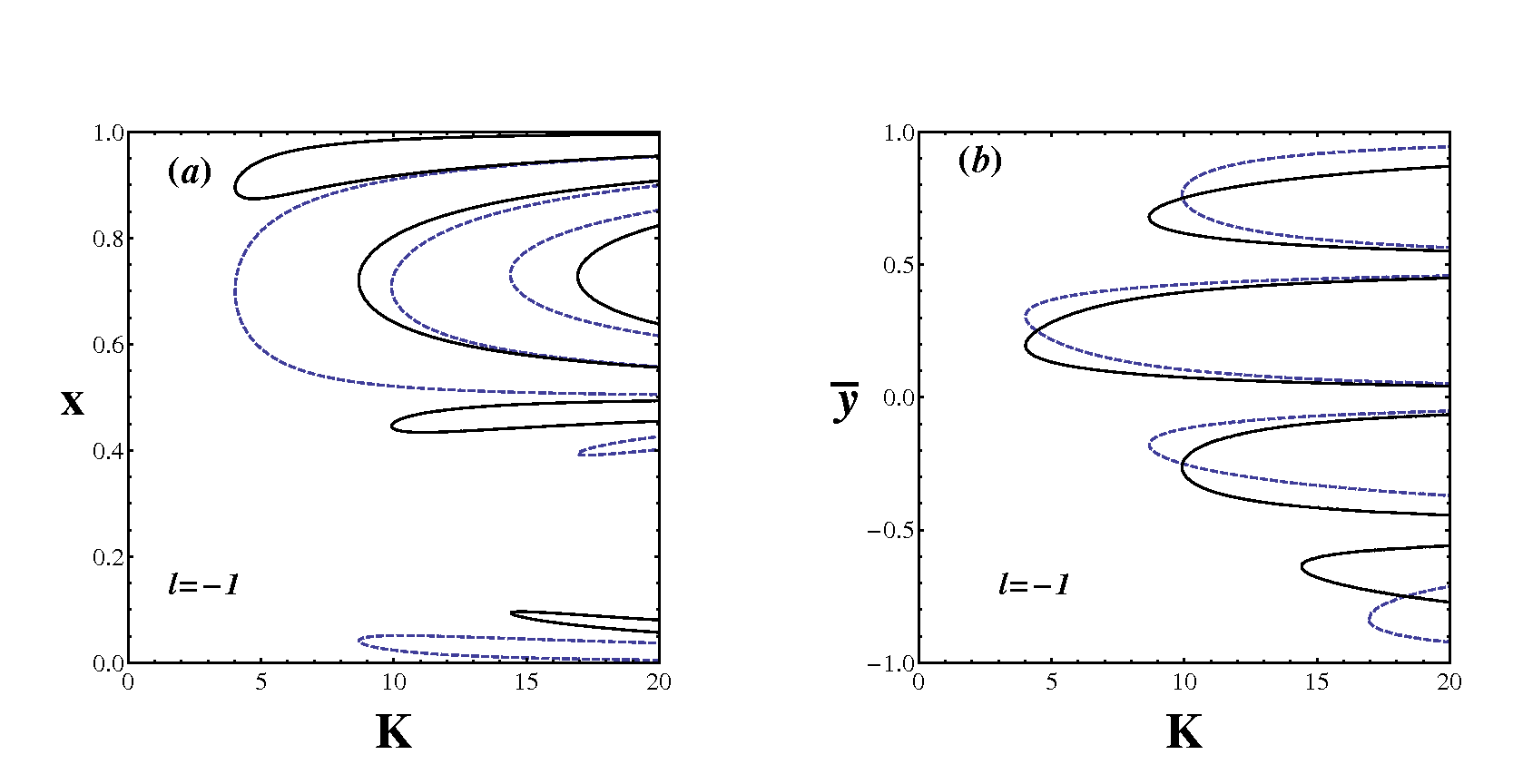}
\caption{(a) Characteristics of accelerator mode periodic orbits of period 2, i.e. $x$ as a function of $K$ derived from equation (\ref{xysinpm}) for $l=-1$ (b) $\overline{y}$ as a function of $K$ derived from eq. (\ref{yx}) for $l=-1$. Black solid curves correspond to equation (\ref{xysinpm})   with the $+$ sign, while (\textbf{blue in online version}) dotted curves correspond to equation  (\ref{xysinpm})  with the $-$ sign. } \label{fig5}
\end{figure*}

\begin{figure*}
\centering
\includegraphics[scale=0.4]{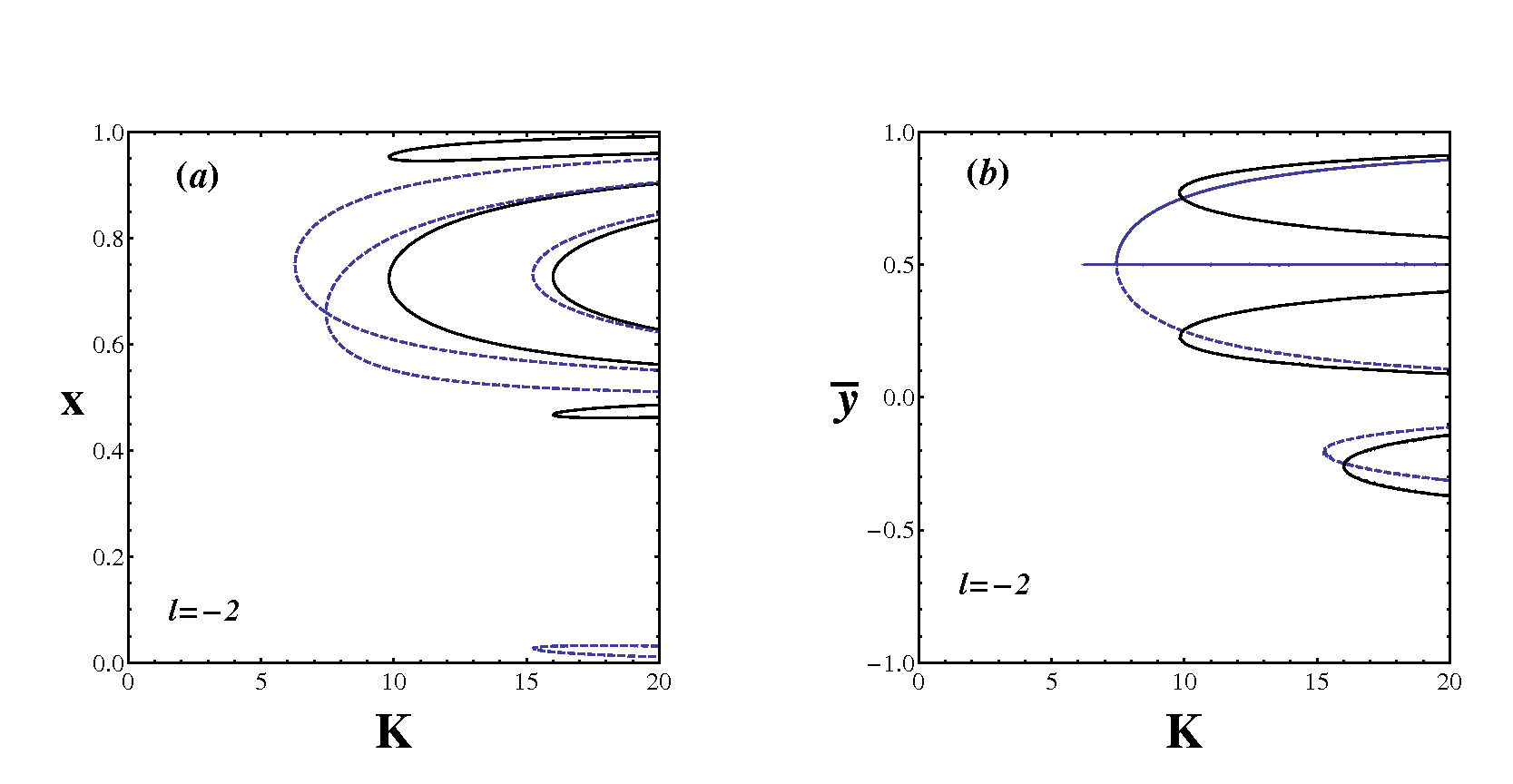}
\caption{(a) Characteristics of accelerator mode periodic orbits of period 2, i.e. $x$ as a function of $K$ derived from equation (\ref{xysinpm}) for $l=-2$ (b) $\overline{y}$ as a function of $K$ derived from eqs. (\ref{yx}) and (\ref{xysinpm}) for $l=-2$. Black solid curves correspond to equation (\ref{xysinpm})    with the $+$ sign, while (\textbf{blue in online version}) dotted curves correspond to equation  (\ref{xysinpm})   with the $-$ sign. (The orbits with constant $\overline{y}=0.5$, i.e. $y=1$, are of period 1.)} \label{fig6}
\end{figure*}

\begin{figure}
\centering
\includegraphics[scale=0.4]{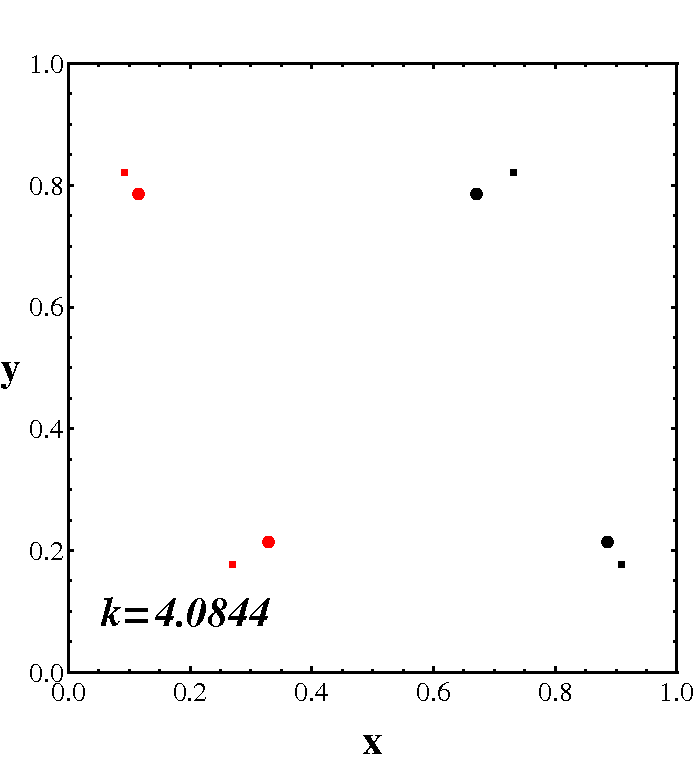}
\caption{Four accelerator double periodic orbits for $K=4.0844$. The two black dots represent a stable periodic orbit with modulo 1 both in $x$ and in $y$ (black squares correspond to the unstable periodic orbit). The two gray (\textbf{red in online version}) dots represent a second stable orbit with $x'=1-x$ and $y'=1-y$ (gray \textbf{(red in online version)} squares correspond to the unstable periodic orbit). } \label{fig7} 
\end{figure}
\begin{figure}
\centering
\includegraphics[scale=0.4]{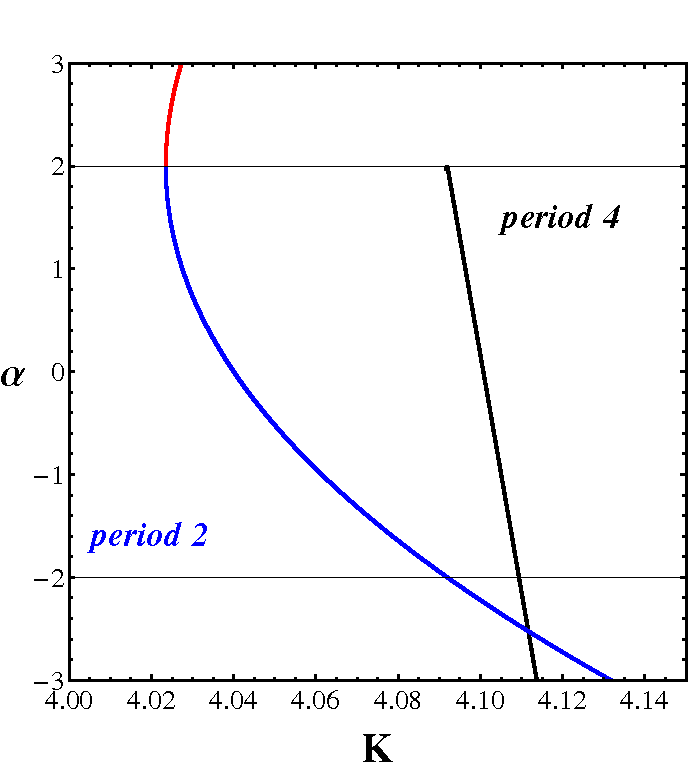}
\caption{The stability diagram of the accelerator mode periodic orbits of Fig. (\ref{fig7}). The orbits are stable for $-2<\alpha<2$. } \label{fig8} 
\end{figure}

The periodic orbits of the groups II and III in \cite{contopoulos2005recurrence} have one initial point along the $x-axis$. However there are further periodic orbits that do not have any point along the $x-axis$ ($y_0$=0). Such orbits are of period $\geq2$.
In this section, we find analytically the characteristic curves of accelerator mode periodic orbits of period 2 that do not have any point on $x$-axis ($y_0\neq0$).  For these periodic orbits, using Eqs. (\ref{stand}) we find:

\begin{equation}\label{yper2}
y_2=y_0+\frac{K}{2\pi} [sin(2\pi x_0)+sin(2\pi x_1)]=y_0+l
\end{equation}
\begin{equation}\label{xper2}
x_2=x_0+2y_0+l+\frac{K}{2\pi}sin(2\pi x_0)=x_0+l'
\end{equation}
where $l$,$ l'$ are integers. For the case of normal modes  $l=0$ while for the case of accelerator modes $l\neq 0$. In the papers of Chiricov (\cite{chirikov1979universal}) and Contopoulos et al. (\cite{contopoulos2005recurrence}) the cases with $l\neq 0$ were mentioned, but they were not explored in detail.  From Eqs. (\ref{yper2}) and(\ref{xper2}) we find:
\begin{equation}\label{yx}
y_0=\frac{l'-l}{2}-\frac{K}{4\pi } sin(2\pi x_0)=\frac{l'-l}{2}-\overline{y_0}
\end{equation} and
\begin{equation}\label{xysin}
sin(2\pi x_1)=sin (2\pi[x_0+\frac{l'-l}{2}+\frac{K}{4 \pi} sin(2 \pi x_0)])
\end{equation}
where $\overline{y_0}=\frac{K}{4 \pi} sin(2\pi x_0)$.
The expression  (\ref{xysin}) is equal to $\pm sin(2 \pi(x_0+ \frac{K}{4 \pi} sin(2 \pi x_0)))$ ($+$ if ($l-l'$) is even and $-$ if ($l'-l$) is odd). Thus from (\ref{yper2}) and (\ref{xysin}) we find:
 \begin{equation}\label{xysinpm}
\frac{K}{2 \pi}[ sin(2 \pi x_0) \pm sin (2 \pi [x_0+ \frac{K}{4 \pi} sin(2 \pi x_0)])]=l
\end{equation}
We solve numerically equations (\ref{yx}) and (\ref{xysinpm}) and we find the characteristics of various families of accelerator periodic orbits of period 2, for various values of $l$. E.g. we find $x_0$ and $y_0$ as functions of $K$ for $l=-1$ in Figs. \ref{fig5}a,b and for $l=-2$ in Figs. \ref{fig6}a,b. Solid black curves correspond to $(l'-l)$=even and dashed blue curves correspond to $(l'-l)=odd$.

Near the minimum values $K=K_{min}$ of Figs. \ref {fig5} and \ref{fig6} there exist stable families, which remain stable  for $K_{min}<K<K_{bif}$, where $K_{bif}$ is the value of $K$ at which  this family becomes unstable, generating, by bifurcation, another family of double period (of multiplicity 4).

An example of two stable accelerator double period orbits near the minimum $K=K_{min} = 4.02$ of Figs. (\ref{fig5}) for $l=-1$ is given in Fig. (\ref{fig7}) for $K=4.0844$. The coordinates of one of the two orbits (black points) are approximately, ($x_0=0.671, ~y_0=0.785$). Then using eqs. (\ref{stand}) we have: ($x_1=0.885, ~y_1=0.215$) and  ($x_2=0.671, ~y_2=-0.215=0.785$ (mod 1)).   Therefore $l'=x_2-x_0=0$ and $l=y_2-y_0=-1$. Thus $l'-l=1$=odd. and $y_0=\overline{y_0} +0.5$.

 The stability diagram of the periodic orbits of Fig. (\ref{fig7}) families is shown in Fig. (\ref{fig8}). The H$\acute{e}$non stability index $\alpha$ denotes a stable orbit whenever it is found inside the interval $-2<\alpha<2$ and unstable elsewhere.  A bifurcation of period-4 orbit is shown as well. This bifurcation is followed by an infinity of period doubling bifurcations and for a little larger $K$ all these families become unstable. 
 
 Similar results are found for $l=-2$, but also for $|l| >2$. Therefore there are many more intervals of $K$ in which accelerator modes exist than those considered in eq. (\ref{accelst}).

In the case $l=-2$ (Fig. \ref{fig6}) the minimum value of $K$ is $K_{min}=2 \pi\approx6.28$. In this particular case the value of  $\overline{y_0}$ is $\overline{y_0}$=0.5 (which is constant with variable $K$) and $l'-l=$odd. Thus $y_0$=1. Using the mapping (\ref{stand}) we find $y_1=0$, $x_1=x_0$, $y_2=-1$, $x_2$=$x_0$-1 etc. i.e. $l'=-1$. Thus these particular orbits are accelerator modes of period 1. All the other orbits of Fig. \ref{fig6} are of period 2.

All these period-2 families are "irregular" (Contopoulos 1970), i.e they have not bifurcated from families existing for $K=0$. In fact  every one of these families is born at a tangent bifurcation, i.e. by joining a stable and an unstable family at a minimum $K=K_{min}$ and exists for all $K>K_{min}$.

\subsection{Higher order accelerator modes }


Higher order stable periodic orbits are generated by bifurcations from the periodic orbits of period 1 or 2. These orbits are called regular, when they are connected to the orbits that exist for $K=0$. However, there are also many irregular  periodic orbits, i.e. periodic orbits that are generated at tangent bifurcations and are independent of the above orbits. Some of these orbits are normal (named group I and group III orbits \cite{contopoulos2005recurrence}) and some orbits are accelerator modes (named group II in \cite{contopoulos2005recurrence}). 

One can find analytically the accelerator periodic orbits of order 4 or higher using the equations of mapping (1). 
For accelerator mode periodic orbits of period-4 with no points on the $x$-axis (i.e. with $y_0\neq0$) the following formulas apply:

\begin{align}\label{or4}
 \nonumber & y_1=y_0+\frac{K}{2 \pi} sin(2 \pi x_0),~~~ x_1=x_0+y_1\\
& y_2=y_1+\frac{K}{2 \pi} sin(2 \pi x_1),~~~x_2=x_1+y_2\\
\nonumber & y_3=y_2+\frac{K}{2 \pi}sin (2 \pi x_2),~~~x_3=x_2+y_3\\
\nonumber & y_4=y_3+\frac{K}{2 \pi} sin(2 \pi x_3)=y_0+\Lambda, ~~~x_4=x_3+y_4=x_0 +\Lambda'
\end{align}
where $\Lambda$ and $\Lambda'$ are integers. In the case of accelerator modes  $\Lambda \neq 0 $ (while $\Lambda'$ may be 0). Using equations (\ref{or4}) we find:

\begin{align}\label{or4acc}
\nonumber &   \frac{K}{2 \pi}[ sin[2 \pi x_0] + 
sin[2 \pi (x_0 + y_0 + \frac{K}{2 \pi} sin[2 \pi x_0])] \\
\nonumber & + sin[2 \pi (x_0 + 2 y_0 + \frac {2K}{2 \pi} sin[2 \pi x_0] \\
\nonumber & +  \frac{K}{2 \pi} sin[2 \pi (x_0 + y_0 + \frac{K}{2 \pi} sin[2 \pi x_0])])]\\
 &  + sin[ 2 \pi (x_0 + 3 y_0 + \frac{3K}{2 \pi} sin[2 \pi x_0]\\
\nonumber &  +   \frac{2K}{\pi} sin[2 \pi (x_0 + y_0 + \frac{K}{2 \pi} sin[2 \pi x_0])]\\
\nonumber & + \frac{K}{2 \pi} sin[2 \pi (x_0 + 2 y_0 + \frac{K}{\pi} sin[2 \pi x_0]\\
 \nonumber &    + \frac{K}{2 \pi} sin[2 \pi (x_0 + y_0 + \frac{K}{2 \pi} sin[2 \pi x_0])])])]]=\Lambda
  \end{align}
and
\begin{align}\label{or4acc2} 
\nonumber & 4 y_0 + \frac{K}{2 \pi}[3 sin[2 \pi x_0]  \\
\nonumber & + 2 sin[2 \pi (x_0 + y_0 + \frac{K}{2 \pi} sin[2 \pi x_0])]\\
& + sin[ 2 \pi (x_0 + 2 y_0 + \frac{2K}{2\pi} sin[2 \pi x_0] \\
\nonumber & + \frac{K}{2 \pi} sin[2 \pi (x_0 + y_0 + \frac{K}{2 \pi} sin[2 \pi x_0])])]]= \Lambda' -\Lambda 
 \end{align} 
  
\begin{figure}
\centering
\includegraphics[scale=0.4]{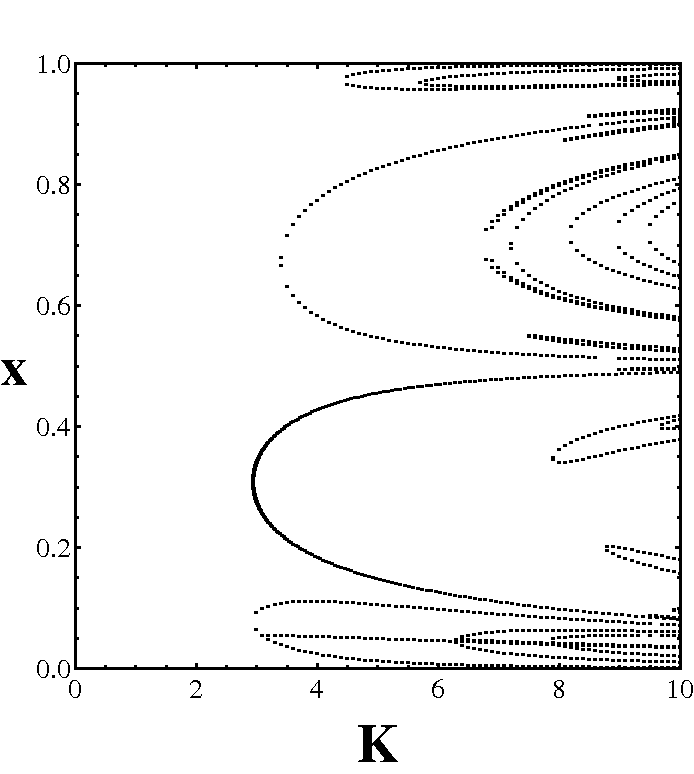}
\caption{Characteristics of accelerator mode periodic orbits of period 4, i.e. $x$ as a function of $K$ derived from equation (\ref{or4acc}) for $\Lambda$=-1. } \label{charper4} 
\end{figure}

For $\Lambda$=0 the orbits are normal, and for $\Lambda \neq $0 the orbits are accelerator modes.   
Using equations (\ref{or4acc}) and (\ref{or4acc2}) and fixed values of $\Lambda$ and $\Lambda'$ we can find the characteristics of the period-4 accelerator mode orbits, i.e. $x_0$ as  function of $K$. 
We give an example for $\Lambda$=-1 and $\Lambda'$=0 in Fig. \ref{charper4} where we plot the characteristics of $x_0$ as a function of $K$ (with the help of "Mathematica"). For values of $K$ near the minimum $K_{min}$ of each curve, we can find stable periodic orbits of period 4. For these values of $K$, the mean value of the diffusion exponent $\mu$, for the whole phase space converges to 2 (after a large enough number of iterations).   

In the same way we find periodic orbits of period 3,5,6,etc.

Duarte in \cite{1994AIHPC..11..359D} has shown that the values of $K$ for which stable periodic orbits exist are dense, i.e. there exist stable periodic orbits in every  interval $\Delta K$, provided $K$ is large enough. We expect that most of these periodic orbits are accelerator modes generating anomalous (ballistic) diffusion. In fact, the cases of normal periodic orbits correspond to $\Lambda$=0, while the cases of accelerator modes correspond to $\Lambda \neq$0, for periodic orbits of any multiplicity, thus they are much more numerous.  However, the sizes of the intervals $\Delta K'$ inside $\Delta K$ where accelerator mode islands exist are very small. In fact, Giorgilli and Lazutkin in \cite{2000PhLA..272..359G} have shown that the total sum of the intervals $\Delta K'$ that contain accelerator mode islands in an interval of $K$ between $n/2$ and $(n+1)/2$ (with $n$ integer) is of order O($1/n$), i.e. quite small if $n$ is large. Thus, we expect that Fig. 1, which gives the global effective diffusion exponent should be supplemented by an infinity of peaks going to infinity (representing anomalous diffusion). Nevertheless, in the great majority of the values of $K$ the global diffusion is normal.  \textbf{An example of a period-4 accelerator mode is given in the Appendix.}

\section{Conclusions}

We studied the global and local diffusion in the standard map. Our main conclusions are the following:\\
(1) We found that in general the global average diffusion of the whole phase space is either normal, when no accelerator mode islands exist (with diffusion exponent $\mu$  equal to 1) or ballistic, when accelerator mode islands exist (and the diffusion exponent is equal to 2). However, in order to find the asymptotic values of $\mu$ a very large number of iterations $n$ is needed (and not simply some thousands or millions) and a large enough number of initial conditions $N$ to ensure population even inside the tiny islands of stability. \\
(2) For values of $K$ where, accelerator modes exist, the diffusion coefficient $D$ for a given (large) number of iterations $n$ is correlated with the total area of the accelerator mode islands.\\
(3) There are a lot of references considering the diffusion as a global process, i.e. over the whole initial phase space ($0<x<1$, $0<y<1$) of the mapping (1) calculating the global mean value of the diffusion exponent $\mu$. However, the diffusion can be different locally depending on the initial conditions in the phase space. We have calculated the local diffusion exponent inside the islands of stability (normal or accelerator), in the sticky region around these islands and in the large chaotic sea for different values of the nonlinearity parameter $K$.\\
 (4) Inside the normal islands of stability the quantity $\langle (y-y_0)^2 \rangle$ is finite and constant with time, thus the diffusion coefficient $D=\langle (y-y_0)^2 \rangle /n$ tends to zero when the number of iterations tends to infinity ($n\rightarrow \infty$). As a consequence, the diffusion exponent $\mu$ tends also to zero. Outside the normal islands of stability the diffusion exponent $\mu$ tends to 1, indicating normal diffusion, in the case where no accelerator mode islands coexist. However, if the initial conditions are in the sticky zone of the normal mode island the diffusion exponent $\mu$  is equal to zero during the time of stickiness and for larger $n$ it converges to 1.\\
(5) Inside the accelerator mode islands the diffusion exponent is $\mu$=2 (ballistic motion). However, outside the accelerator mode islands the diffusion in general is normal, i.e. $\mu \rightarrow 1$, when $n\rightarrow \infty$. Initial conditions in the chaotic sea far away from the sticky zone will finally get diffused normally with $\mu \rightarrow 1$  for $n\rightarrow \infty$. 
Initial conditions inside the sticky zone are dragged by the accelerator mode island, but only for a specific time interval. This dragging time depends on the distance from the island of stability. In fact, the dragging time is roughly equal to the escape time (or initial stickiness time), i.e the time required by an orbit to escape to the large chaotic sea far from the islands of stability. The orbits spend a long time in the chaotic sea before entering again into the sticky zone and the Poincar$\acute{e}$ recurrence time is much larger than the stickiness time. Moreover, the Poincar$\acute{e}$ recurrence time is different for nearby initial conditions in the sticky zone, because these orbits deviate considerably after a Lyapunov time (which is much shorter than the Poincar$\acute{e}$ recurrence time). Thus, although individual orbits enter again in the sticky zone and are again dragged by the accelerator modes, they do not come back in the sticky zone simultaneously. In fact, the distribution of the points of different orbits becomes uniform in the chaotic sea and in the sticky zone, as times goes to infinity. So, even if we take initial conditions in the extreme sticky region, these chaotic orbits will finally manifest a normal diffusion.  \\
(6) The stable accelerator periodic orbits become unstable as the nonlinearity parameter $K$ becomes larger and bifurcate \textbf{into} orbits of larger multiplicity. These  unstable accelerator mode periodic orbits exist for all values of $K$ but they do not generate any lasting anomalous diffusion around them.\\
(7) Normal and accelerator mode islands coexist for some intervals of the nonlinearity parameter $K$. Then, the global diffusion exponent is $\mu=2$. The local diffusion exponent inside the normal islands is $\mu=0$ and inside the accelerator mode islands is $\mu=2$, while outside the islands of stability, the local diffusion exponent converges slowly to $\mu=1$ after a long enough time.\\
(8) We have found analytically the characteristic curves of several cases of accelerator modes of period 1,2 and 4 that whenever they are stable generate anomalous diffusion.\\
(9) Between any two nearby values of $K$, which differ by a small $\Delta K$, there always exist \textbf{values of $K$ for which there exist} islands of stability. However, the measure of these intervals is small. We expect that most of these islands are accelerator modes, because as we have shown in subsection \rom{4}.B, the cases of accelerator modes are much more numerous than \textbf{those for} the normal modes. Thus, we expect that the values of $K$ for which the global diffusion exponent is equal to $\mu$=2 are dense. Nevertheless, for the majority of the values of $K$ the global diffusion is normal, with $\mu$=1.

\appendix*

\section{An example of a period-4 accelerator mode}

 A particular case of period-4 orbits has the following properties:

\begin{align}\label{diff5}
& y_1= 1-y_0\\
&  y_3=-y_2 
\end{align}

Then
\begin{align} \label{diff6}
\nonumber & x_1= 1+x_0-y_0\\
 & y_2=y_1+\frac{K}{2\pi}\sin(2\pi x_1),~~~ x_2=1+x_0-y_0+y_2\\
\nonumber & y_3=y_2+\frac{K}{2\pi}\sin(2\pi x_2)=-y_2,~~~x_3=1+x_0-y_0=x_1
\nonumber
\end{align}
Thus we derive
\begin{eqnarray} \label{diff7}
y_4=y_0-1, ~~~x_4= x_0\\
\nonumber
\end{eqnarray}
therefore, this is an accelerator mode periodic orbit.
From Eqs. (\ref{diff6})-(\ref{diff7}) we find
two equations relating $K$, $x_0$ and $y_0$:

\begin{align}\label{diff8}
\frac{K}{2\pi} \sin(2\pi x_0)=1-2y_0
\end{align}
and
\begin{align}\label{diff9}
\nonumber & \frac{K}{2\pi}\sin
2\pi[(1+x_0-y_0)+(1-y_0+\frac{K}{2\pi}\sin 2\pi (1+x_0-y_0)
)]\\ &
 =-2y_2=-2+2 y_0- \frac{2K}{2\pi} \sin (2\pi(1+x_0-y_0))
\end{align}

\begin{figure*}
\centering
\includegraphics[scale=0.4]{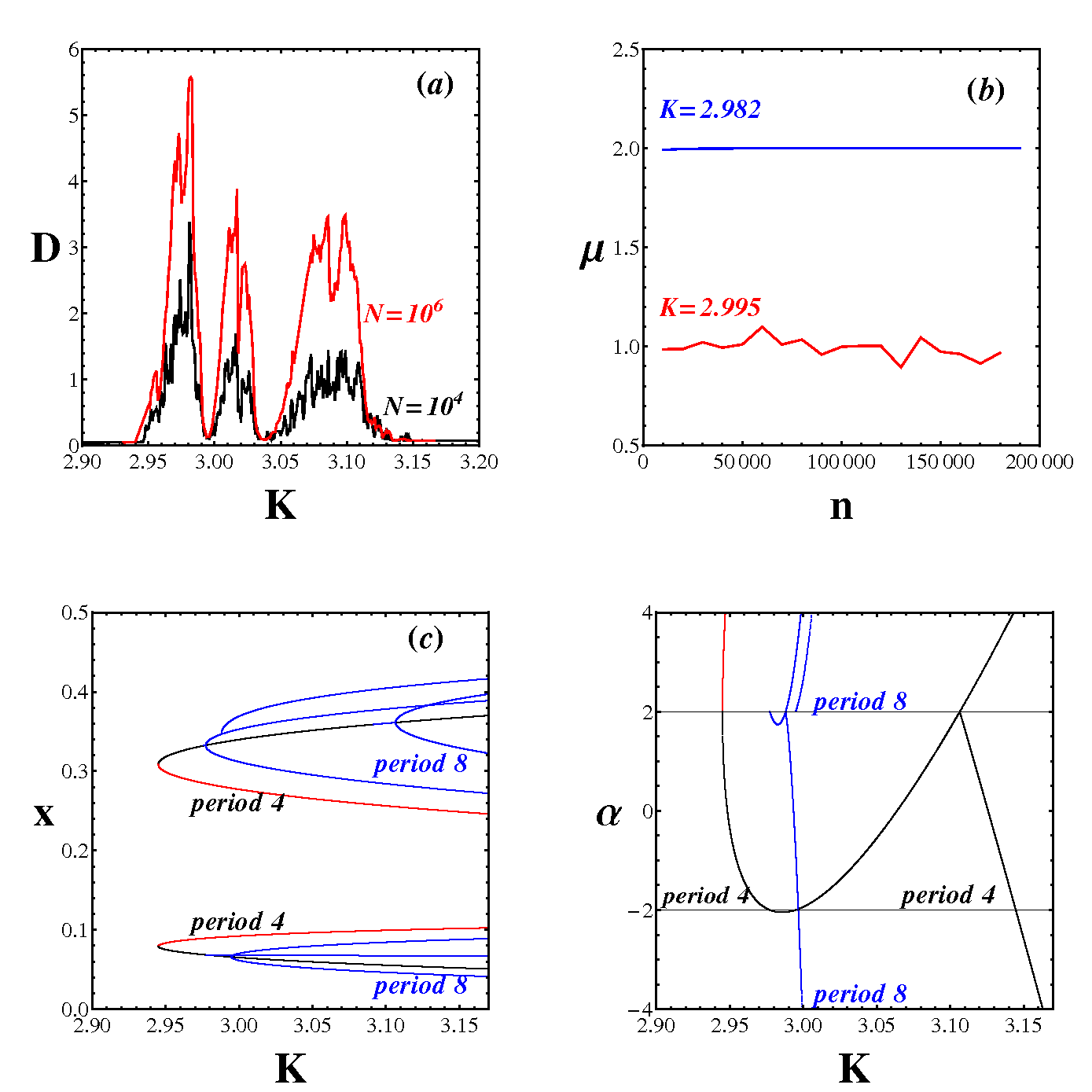}
\caption{(a) The diffusion coefficient $D(K)$ of accelerator modes for values of
the non linearity parameter $2.9<K<3.2$. There are intervals of $K$
in which the diffusion coefficient $D$ increases with the number
of iterations $n$ and small intervals that seem to stay constant with the number of iterations (but they still correspond to anomalous diffusion, see text). The black curve is derived from a grid of $N=10^5$ initial conditions, while the \textbf{gray (red in online version)} curve is derived from a grid of $N=10^6$ initial conditions.  (b) The time evolution of the global diffusion exponent $\mu$ as a function of
the number of iterations $n$ for two nearby values of $K$. For $K=2.982$ the diffusion exponent is $\mu=2$ denoting anomalous diffusion but for $K=2.995$ it has a mean value around $\mu=1$ denoting normal diffusion. However, this result is not true for the case of $K=2.995$ and it is due only to the small number of initial conditions taken in the whole phase space that do not populate the small accelerator mode islands of stability (see text). (c) Accelerator mode periodic orbits of period 4 (black, initially stable and \textbf{gray (red in online version)}, unstable) and period 8 (\textbf{blue in online version)} existing in the same interval as in (a). (d) The stability index $\alpha$ of the accelerator mode periodic orbits of (c).} \label{difmk3}
\end{figure*}

Using  Eqs. (\ref{diff8}) and (\ref{diff9}) we find  an example of stable accelerator mode periodic orbit of period 4, for $K=2.97$, with $x_0=0.06988$, $y_0=0.39983$. The points
($x_3,y_3$) and ($x_4,y_4$) are outside the square (0,1)x(0,1). But
if we use modulo 1 we find $x'_3=x_1$, $y'_3=1-y_3$ and $x'_4=x_0$,
$y'_4=y_0$. The higher order images of the point ($x_0,y_0$) are further
outside the square $(0,1) \times (0,1)$ thus the orbit forms an
accelerator mode. Similar results are found for 4 more islands with $\tilde{x_0}=1-x_0$ and $\tilde{y_0}=1-y_0$. This stable periodic orbit is created at $K\approx 2.945$ at a tangent bifurcation together with an unstable periodic orbit (Fig. \ref{difmk3}). 

 The H$\acute{e}$non stability parameter $\alpha$ of
this orbit is given in Fig. (\ref{difmk3})d. The curve $\alpha (K)$ crosses
the axis $\alpha =-2$ at two nearby points (near $K=2.98$). At these points it generates two families of
period-8 (blue curves in Figs. \ref{difmk3}c,d). The first period-8 family has a stability index $\alpha=2$ when it is created and it is initially stable but it crosses the axis $\alpha =2$ again and generates another period-8 family.  The second period-8 family is  unstable. The stability parameter $\alpha$ of the period-4 family goes beyond $\alpha=2$ for $K\approx 3.14$, generating by
bifurcation another stable family of period-4 which is followed
by a cascade of period-doubling bifurcations. The corresponding 
diffusion coefficient for these accelerator mode families is shown in Fig. (\ref{difmk3})a after a number of
$n=10^4$ iterations (black curve corresponds to a grid of $N=10^4$ initial conditions and red curve corresponds to a grid of $N=10^6$ initial conditions).  As the number of iterations increases the
diffusion coefficient $D$ increases and the whole region of $K$ with $2.95<K<3.15$ is characterised by anomalous diffusion.
Nevertheless, there are small intervals of $K$ (near $K\approx 2.99$ and $K\approx 3.04$) that seem to
correspond to normal diffusion, i.e. the value of the 
diffusion coefficient seems to stay close to a small constant value, even when we use a much larger number of initial conditions ($N=10^6$, in red curve). We have investigated these two cases in detail.

\begin{figure*}
\centering
\includegraphics[scale=0.35]{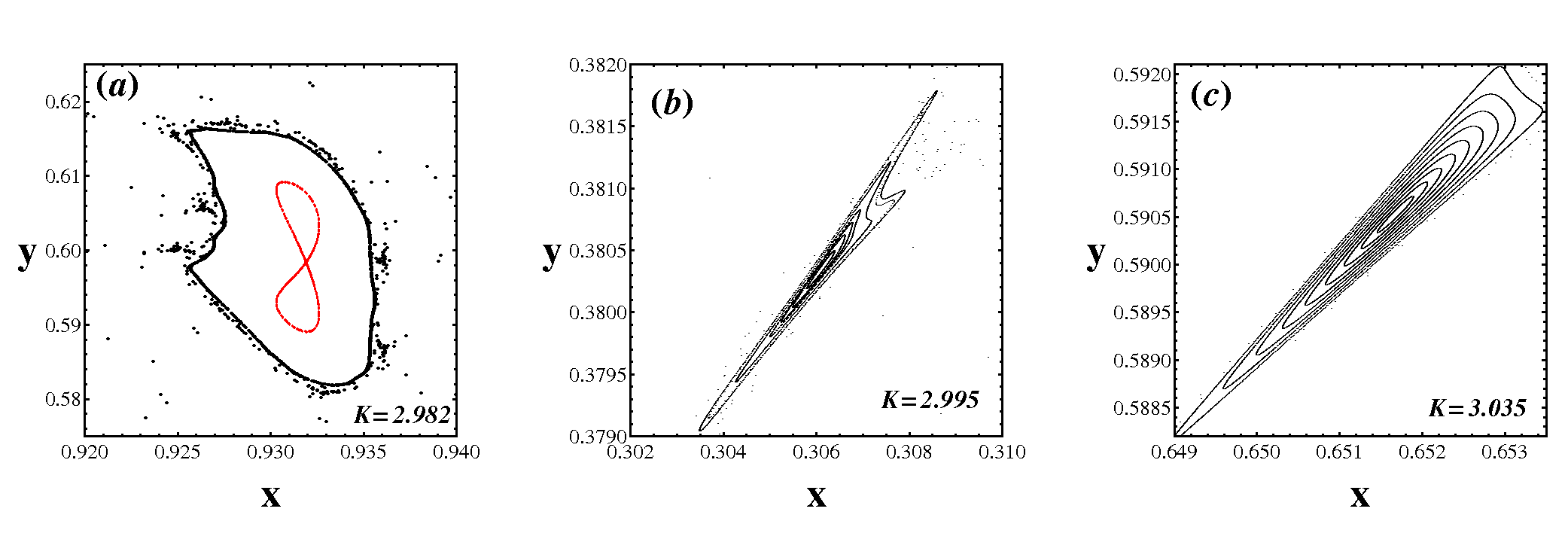}
\caption{(a) For $K=2.987$ the accelerator mode unstable periodic orbit of period 4 is inside the last KAM curve (black curve) although this orbit is unstable. Its asymptotic curves (\textbf{gray (red in online version)}) surround two points representing a bifurcated periodic orbit of period 8.  (b) For $K=2.995$ the period-4 orbit is unstable, but there is a stable periodic orbit of period-8 that has bifurcated from the period-4 orbit, which is surrounded by an island (one of the 8 islands). (c) The accelerator mode stable epriodic orbit of period 4 (one of 4 islands of stability) for $K=3.305$. In all three cases the area of the periodic orbits corresponds to anomalous diffusion and gives a global diffusion exponent $\mu=2$. } \label{difmk4}
\end{figure*}

 We have calculated the global diffusion exponent $\mu$ for two nearby values of $K$ in Fig. \ref{difmk3}b, for the whole phase space and for a number $n=10^4$ of iterations  and $5 \times 10^4$ initial conditions. For $K=2.982$ the mean value of the diffusion exponent $\mu$ is 2, which denotes anomalous diffusion.
However, for $K=2.995$ the value of $\mu$ stays for a long time close to $\mu=1$, and this indicates normal diffusion. In order to give an explanation for these minima of Fig.\ref{difmk3}a we studied separately the phase space for three values of $K$, namely $K=2.982$ (where there is a maximum of the diffusion coefficient value $D$) and $K$=2.995, $K$=3.035 (where there are  minima of the diffusion coefficient $D$). In Fig. \ref{difmk4}a we plot  a part of the phase space for $K=2.982$. The period-4 accelerator mode periodic orbit has become unstable and a period-8 stable periodic orbit has bifurcated, but there still exist a KAM curve surrounding the unstable periodic orbits of period-4 and period-8. The area inside this last KAM curve corresponds to anomalous diffusion and therefore the whole phase space has a mean diffusion exponent $\mu=2$ (see Fig. \ref{difmk3}b.)  In Fig. \ref{difmk4}b we plot  a part of the phase space for $K=2.995$ where a stable periodic orbit of period-8 is shown. Even if the area of this island of stability (one of 8 islands of stability) is tiny, for large enough time and large enough number of initial conditions inside the whole phase space, the anomalous diffusion of this accelerator mode island will dominate the whole phase space and the diffusion coefficient $D$ should increase with the number of iterations $n$ and the corresponding diffusion exponent $\mu$ should finally converge to $\mu=2$. Therefore, the minimum of the red curve shown in Fig. \ref{difmk3}a in $K=2.995$ is only due to the insufficient number of initial conditions ($10^6$) that do not populate the area of the tiny accelerator mode islands. 
 Finally in Fig. \ref{difmk4}c we plot  a part of the phase space for $K=3.035$ where a stable periodic orbit of period-4 is shown. Here again the minimum in  Fig. \ref{difmk3}a for this value of $K$ is due only to the insufficient number of initial conditions.

The conclusion is that in the whole range of the values of $K$ between $K=2.95$ and $K=3.15$ shown in Fig. \ref{difmk3}a there exist stable periodic orbits of accelerator mode and islands around them therefore the whole phase space should present a mean value of the diffusion exponent $\mu$ which converges to $\mu=2$.

\bibliographystyle{plain}
\bibliography{poincare-new}

\end{document}